\documentclass[aps,prd,twocolumn,amsmath,superscriptaddress,amssymb,showpacs,floatfix,nofootinbib,longbibliography]{revtex4-1}
\pdfoutput=1
\usepackage{graphicx}
\usepackage{bm}
\usepackage{times}
\usepackage[dvipsnames]{xcolor}
\usepackage{hyperref}
\hypersetup{
    colorlinks = true,
    linkcolor = {MidnightBlue}, 
    citecolor = {MidnightBlue},
    urlcolor = {MidnightBlue},
}
\usepackage{slashed}
\usepackage{color}
\usepackage{aas_macros}
\usepackage{slashed}
\usepackage{lipsum}
\usepackage{subfigure}
\usepackage{multirow}
\usepackage{amsmath}
\usepackage{array} 
\usepackage{varwidth} 

\usepackage{xr}

\newcommand{\be}{\begin{equation}}
\newcommand{\ee}{\end{equation}}
\newcommand{\bea}{\begin{eqnarray}}
\newcommand{\eea}{\end{eqnarray}}

\begin{document}

\title{Pulsars Do Not Produce Sharp Features in the Cosmic-Ray Electron and Positron Spectra}
\author{Isabelle John}
\email{isabelle.john@fysik.su.se, ORCID: orcid.org/0000-0003-2550-7038}
\affiliation{Stockholm University and The Oskar Klein Centre for Cosmoparticle Physics,  Alba Nova, 10691 Stockholm, Sweden}
\author{Tim Linden}
\email{linden@fysik.su.se, ORCID: orcid.org/0000-0001-9888-0971}
\affiliation{Stockholm University and The Oskar Klein Centre for Cosmoparticle Physics,  Alba Nova, 10691 Stockholm, Sweden}
\begin{abstract}
Pulsars are considered to be the leading explanation for the excess in cosmic-ray positrons. A notable feature of standard pulsar models is the sharp spectral cutoff produced by the increasingly efficient cooling of very-high-energy electrons by synchrotron and inverse-Compton processes.
This spectral break has been used to argue that many pulsars contribute to the positron flux and that spectral features cannot distinguish between dark matter and pulsar models. We prove that this feature does not exist --- it appears due to approximations that treat inverse-Compton scattering as a continuous, instead of as a discrete and catastrophic, energy-loss process. Astrophysical sources do not produce sharp spectral features via cooling, reopening the possibility that such a feature would provide evidence for dark matter.
\end{abstract}

\maketitle

\section{Introduction}
Observations by PAMELA~\cite{2010PhRvL.105l1101A} and AMS-02~\cite{PhysRevLett.113.121101, AMS:2019iwo, AMS:2019rhg} have provided clear evidence for a rise in the positron fraction at energies above $\sim$10~GeV. This excess has most commonly been interpreted as either evidence of dark matter (e.g.,~\cite{Arkani-Hamed:2008hhe}) or the production of electron and positron pairs (e$^\pm$, hereafter, electrons) by energetic pulsars~\cite{Hooper:2008kg, Profumo:2008ms}. Over the last five years, TeV halo observations have shown that pulsars efficiently convert a large fraction of their spin-down power into energetic electrons, providing credence to the pulsar explanation~\cite{Hooper:2017gtd, HAWC:2017kbo, Profumo:2018fmz, Martin:2022aun, Liu:2022hqf}. TeV halo observations also have intriguing effects for our understanding of diffusion throughout the Milky Way~\cite{Lopez-Coto:2022igd}.

In addition to energetic arguments, the positron spectrum has long been discussed (even before PAMELA) as a discriminant of the underlying mechanism. Dark matter models generically include sharp spectral ``lines" at an energy corresponding to the dark matter mass~\cite{Turner:1989kg}. However, pulsar models include their own sharp spectral feature located at an energy determined by the pulsar age and the energy-loss rate of very high-energy electrons~\cite{1995A&A...294L..41A}. These features are similar, and models for the PAMELA and AMS-02 data have discussed the difficulty in using spectral features to constrain their dark matter or pulsar origin~\cite{Malyshev:2009tw, Barger:2009yt, 2010ApJ...710..958K, Linden:2013mqa}, a topic which was revived after DAMPE observations~\cite{DAMPE:2017fbg} of a potential 1.4~TeV electron spectral bump~\cite{Wang:2017hsu, 2018ApJ...854...57F, Huang:2017egk, Fornieri:2019ddi, Bao:2020ila}. 

As $\gamma$-ray data have begun to prefer the pulsar interpretation~\cite{Hooper:2017gtd, Fang:2018qco, Profumo:2018fmz}, studies have focused on whether the excess is dominated by a few nearby pulsars or a large ensemble of systems~\cite{Hooper:2008kg, Profumo:2008ms, Cholis:2013psa, Asano_2022, Linden:2013mqa, Hooper:2017gtd}. Spectral considerations again play an important role. Models predict that every pulsar will produce a spectral cutoff at an energy corresponding to the pulsar age. The detectability of this feature depends on its fractional contribution to the positron flux. Because AMS-02 does not find any sharp spectral features, models tend to prefer scenarios where many pulsars contribute to the excess~\cite{Cholis:2017ccs, Fang:2017nww, Cholis:2018izy, Orusa:2021tts, Cholis:2021kqk}.

\begin{figure}[tbp]
\centering
\includegraphics[width=0.47\textwidth]{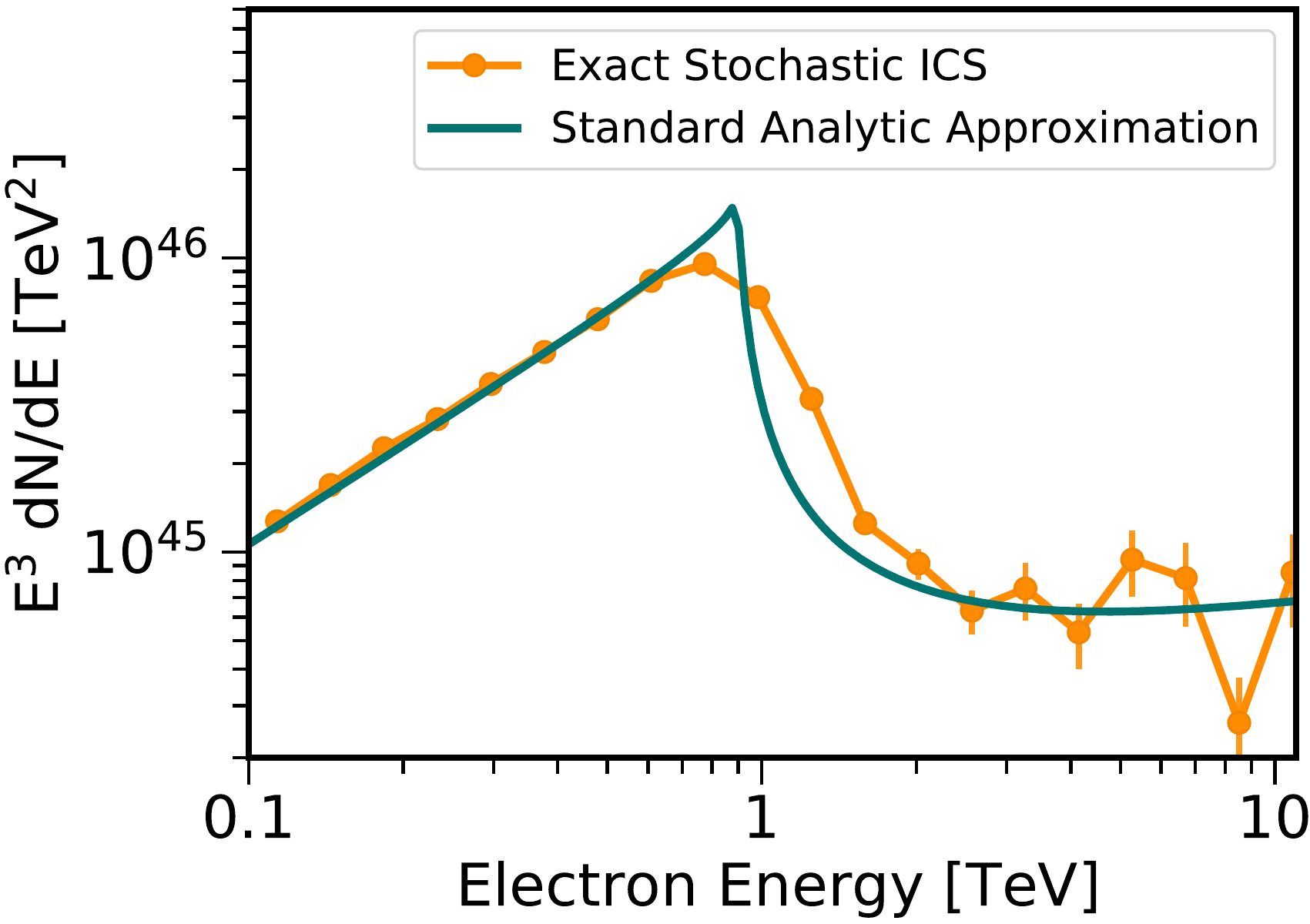}
\caption{
The total electron spectrum integrated over all distances from a 342~kyr old pulsar ({\it e.g.,} Geminga), obtained by the standard analytic approximation (blue), compared to our exact stochastic ICS model (orange). The sharp spectral feature at $\sim 0.88$~TeV is washed out, while the low and high energy behavior remains the same. Error bars are statistical due to the low-number of high-energy counts.}
\label{fig:1}
\end{figure}

The reason for this spectral feature is straightforward. Young pulsars spin down quickly, injecting most of their electrons in a few thousand years. These electrons cool rapidly through synchrotron and inverse-Compton scattering (ICS). Critically, both processes cool electrons at a rate that is proportional to the square of the electron energy. Thus, the highest energy electrons all cool to almost the same critical energy regardless of their initial energy. Because the electrons are born at about the same time and travel through the same magnetic and interstellar radiation fields, they bunch up at a specific energy, above which there is a sharp cutoff.

However, this explanation depends on an incorrect simplification. It assumes that ICS is a continuous process where numerous interactions each remove infinitesimal energy from the electron. Instead, the ICS of high-energy electrons is a catastrophic process, where individual interactions remove a large fraction ($\sim$10--100\%) of the electron energy.

In this paper, we prove that when the stochasticity of ICS is correctly modeled, pulsars do not produce a sharp feature in the local cosmic-ray spectrum. Using a detailed Monte Carlo, we find that ICS energy losses typically produce a distribution of electron energies that are dispersed by $\sim$50\% around the standard ``critical value", washing out the sharp spectral effect (see Figure~\ref{fig:1}). Importantly, this result does not apply to dark matter models, where the spectral feature is instead directly produced in the annihilation event. Thus, our results have significant implications for our ability to differentiate dark matter and pulsar contributions to the positron flux. 

\section{Modeling}
\subsection{Standard Pulsar Models} 
Most pulsar studies use similar approaches, which we detail in Appendix~\ref{app: standardmodel}. Here, we review \emph{three} points that are relevant for our results. The \emph{first} is that most electrons are accelerated when the pulsar is young. The electron luminosity traces the pulsar spindown power as:

\begin{equation}
\label{eq:luminosityvstime}
    L(t) = \eta L_0 \left ( 1+\frac{t}{\tau} \right )^{-2}
\end{equation}

\noindent where $L_0$ is the initial luminosity, $\eta$ is a conversion efficiency, and $\tau$ is a timescale, which theory and data show to be $\mathcal{O}$(10~kyr). This is short compared to the electron diffusion and cooling timescales, meaning that pulsars inject a significant fraction of their total electron energy before the electrons cool considerably.

The \emph{second} point is that the electron injection spectrum is hard and continues to very high energies. Standard models use an injection spectrum:

\vspace{-0.2cm}
\begin{equation}
\label{eq:pulsarspectrum}
    \frac{dN}{dE} =  Q(t) ~ E^{-\alpha} e^{-E/E_{cut}}
\end{equation}

\noindent where $Q(t)$ is a normalization related to $L(t)$. Best-fit values for $\alpha$ span 1.5 -- 2.2~\cite{Hooper:2017gtd, HAWC:2017kbo, DiMauro:2019yvh}, with E$_{cut}$ between 0.01--1~PeV, though see Refs.~\cite{Caprioli:2008sr, Bucciantini:2018dkr} for more detailed models. This means that most of the electron power is injected far above the GeV-scales where the local electron flux is best-measured.

The \emph{third} point is that the energy-loss rate for high-energy electrons gets faster at higher energies, with a form: 

\begin{equation}
\label{eq:standardenergylosses}
    \frac{dE}{dt} = -\frac{4}{3}\sigma_T c \left( \frac{E}{m_e}\right)^2 \left[ \rho_{B} + \sum_i \rho_i(\nu_i) S(E,\nu_i ) \right ]
\end{equation}

\noindent where $\sigma_T$ is the Thomson cross section, $E$ and $m_e$ are the electron energy and mass, $\rho_B$ is the magnetic field energy density, and $\rho_i$ are the energy densities of interstellar radiation field (ISRF) components with energies $\nu_i$. \mbox{S$_i$(E, $\nu_i$)} accounts for the Klein-Nishina suppression of the ICS cross-section~\cite{Blumenthal:1970gc}. For a magnetic field \mbox{$B\sim$~3~$\mu$G = 0.22~eV~cm$^{-3}$ } and ISRF of 1~eV~cm$^{-3}$ this timescale is:

\begin{equation}
\label{eq:energylosstimescale}
    t_{loss} \approx 320~{\rm kyr} \left( \frac{E}{1~{\rm TeV}}\right)^{-1}\left(\frac{\rho_{{\rm tot}}~S_{{\rm eff}}(E)}{1~{\rm eV}~{\rm cm}^{-3}} \right)
\end{equation}

\noindent where $S_{{\rm eff}}$ is calculated by convolving the Klein-Nishina effect from each ISRF component. From this, we see that 1~TeV (100 TeV) electrons require $\sim$300~kyr (only $\sim$3~kyr) to cool. Finally, these terms are integrated into a diffusion equation, which determines the time-dependent electron flux that propagates from a pulsar to Earth.

Eqns.~(\ref{eq:luminosityvstime} -- \ref{eq:energylosstimescale}) show how pulsars produce a spectral feature. The pulsar produces very-high-energy electrons in a $\delta$-function-like burst. The most energetic electrons cool more quickly than lower-energy electrons, producing a spectrum that ``bunches up" at a critical energy, $E_c$. This critical energy decreases with pulsar age, but is between 100--1000~GeV for the pulsars that are most critical to the AMS-02 observations. This result is generic to any pulsar model that uses Eqn.~(\ref{eq:standardenergylosses}) to calculate ICS cooling.  \\

\subsection{The Stochasticity of Inverse-Compton Scattering} The problem with this approach stems from Eqn.~(\ref{eq:standardenergylosses}), which calculates the \emph{average} energy that an electron loses over a period of time, but does not account for the \emph{dispersion} in these losses. Eqn.~(\ref{eq:standardenergylosses}) treats energy losses as continuous, when they, in fact, stem from a finite number of interactions between an electron and ambient magnetic and radiation fields.

For synchrotron radiation, the difference is negligible. The critical energy for synchrotron radiation is given by: 

\begin{equation}
    \nu_c = \frac{3\gamma^2eB}{4\pi m_e c} = 0.06 \left(\frac{B}{1~{\rm \mu G}}\right)\left( \frac{E}{1~{\rm TeV}} \right)^2~{\rm eV}
\end{equation}

\noindent Thus, the energy loss from each interaction is small, as is the relative variance in the number of interactions (1/$\sqrt{n}$).

For ICS, however, individual interactions are important. The ICS differential cross-section was originally computed in Ref.~\cite{KN:1928, Blumenthal:1970gc}, and is reported here from Ref~\cite{1981Ap&SS..79..321A}:

\begin{multline}
    \label{eq:fullkn}
    \frac{d^2\sigma(E_\gamma, \theta)}{d\Omega dE_\gamma} = \frac{r_0^2}{2\nu_iE^2} ~ \times ~ \\ \left [1 +   \frac{z^2}{2(1-z)} - \frac{2z}{b_\theta(1-z)} +  \frac{2z^2}{b_\theta^2(1-z)^2}\right ]
\end{multline}

\noindent where $E_\gamma$ is the final $\gamma$-ray energy, $\nu_i$ and $E$ are the initial energies of the photon and electron and $\theta$ is the angle between them, and $r_0$ is the classical electron radius. The parameter \mbox{$z$ $\equiv$ $E_\gamma$/$E$} and \mbox{$b_\theta$ $\equiv$ 2 (1-cos $\theta$)$\nu_iE$.} At low energies (Thomson regime), only the first term is non-zero, the cross-section is $\sigma_T$~=~6.6 $\times$ 10$^{-25}$~cm$^2$, and the relevant energy scales for this process are approximately:

\begin{equation}
    E_{\gamma,c} = \frac{4}{3} \gamma^2 \nu_i
\end{equation}

At high energies (Klein-Nishina regime) the critical energy, \textit{i.e.}, the average energy lost, exceeds the electron energy (which is kinematically forbidden), suppressing the cross-section and producing $\gamma$-rays with energies just below the electron energy. 

Thus, even for scatterings with typical CMB photons ($\nu_i$~$\sim$~10$^{-3}$~eV), an electron with an initial energy of 1~TeV loses 5~GeV, a 10~TeV electron loses 500~GeV, and a 100~TeV electron loses 50~TeV. Energy losses for infrared ($\nu_i$~$\sim$ 10$^{-2}$~eV), optical ($\nu_i$~$\sim$~1~eV), and UV ($\nu_i$~$\sim$~10~eV), are even higher, and can fall well into the Klein-Nishina range. \\

\subsection{Interstellar Radiation Field Model}
We model the interstellar radiation field (ISRF) based on four components: the cosmic-microwave background (energy density \mbox{$u=0.26$~eV/cm$^3$}, temperature \mbox{$T=2.7$~K}), infrared (\mbox{$u=0.60$~eV/cm$^3$}, \mbox{$T=20$~K}), optical (\mbox{$u=0.60$~eV/cm$^3$}, \mbox{$T=5000$~K}) and ultra-violet (\mbox{$u=0.10$~eV/cm$^3$}, \mbox{$T=20000$~K}) \cite{Hooper:2017gtd}. From this, we compute the photon number density in 560 logarithmic bins spanning from \mbox{10$^{-5}$ -- 200~eV} following a blackbody spectrum, and use Monte Carlo techniques to select target photons from this distribution.

\subsection{Geminga as a Template}
In our default analysis, we choose model parameters consistent with values for Geminga, a nearby ($\sim$250~pc), middle-aged \mbox{($\sim$ 342~kyr)} pulsar~\cite{Manchester:2004bp}. We set the electron injection index $\alpha = 1.9$ and energy cutoff $E_\text{cut} = 100$~TeV~\cite{Hooper:2017gtd}, normalized to the total energy output of Geminga $E_\text{tot} \approx 9.8\times10^{50}$~GeV, and an efficiency of converting spindown power into $e^\pm$ pairs of $\eta = 0.1$. We adopt a time-dependent luminosity following Eqn.~(\ref{eq:luminosityvstime}), with a spin-down timescale of 9.1~kyr~\cite{Hooper:2017gtd}. This last parameter may vary significantly between pulsars~\cite{Fornieri:2019ddi, Orusa:2021tts}. We note that our analysis and results apply to any young and middle-aged pulsar, and only choose Geminga as an example here. 

\subsection{Numerical Setup}
We use a Monte Carlo approach to account for the variance of ICS energy losses. The electron energy is calculated explicitly in time as follows: (1)
we begin with an electron formed at a time $t$ after pulsar formation and initial energy $E_0$, (2) we evolve the system in time, choosing a time step small enough that synchrotron losses (assuming a magnetic field strength of 3~$\mu$G) and the probability of having two ICS events are negligible, (3) based on the electron energy and photon density, we randomly pick whether an ICS event happens or not, and if so calculate the initial and final photon energy, (4) we re-compute the electron energy and repeat this process up to the current pulsar age.

\begin{figure}[tbp]
\centering
\includegraphics[width=0.49\textwidth]{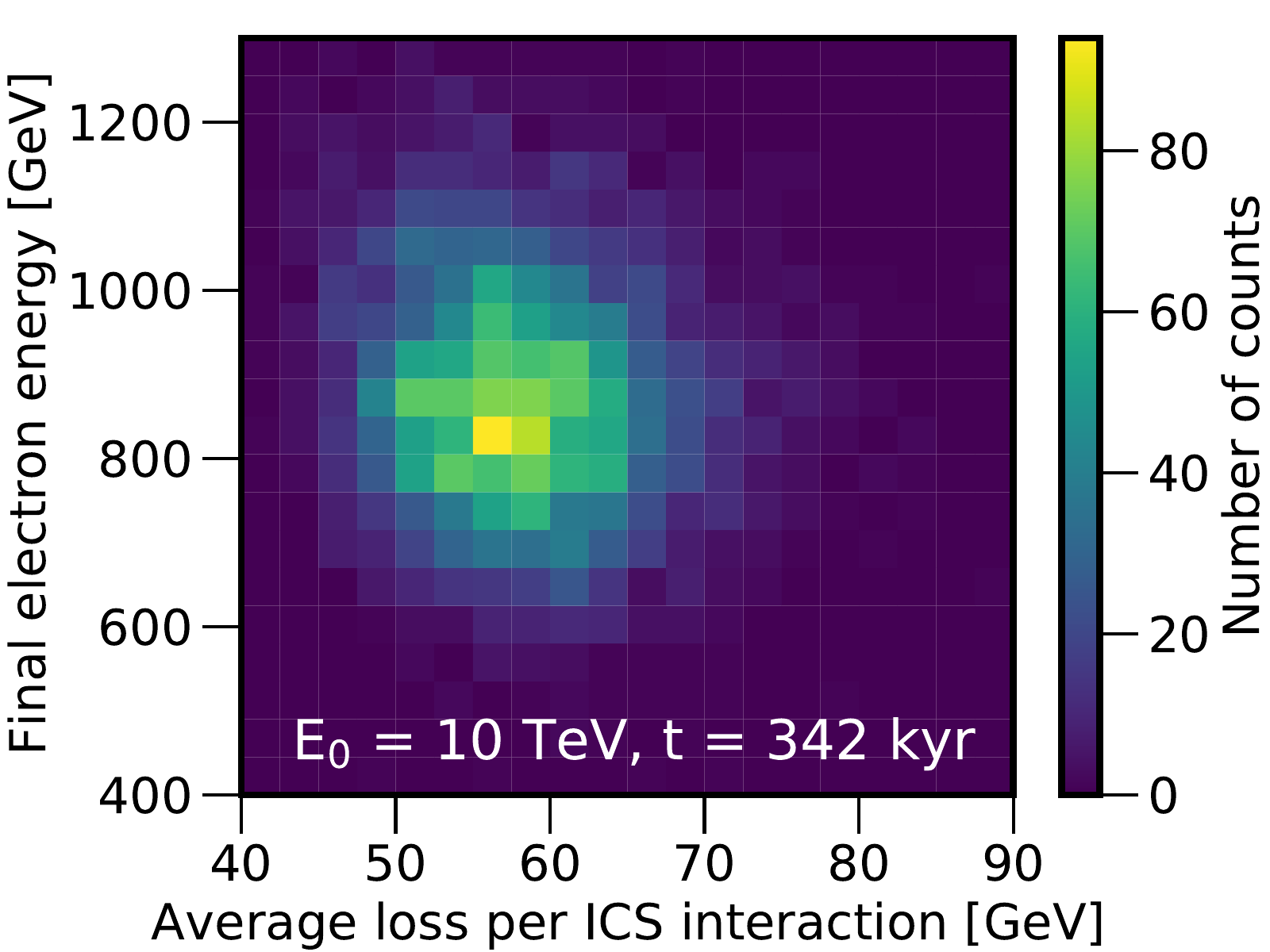}
\caption{The electron energy after 342~kyr compared to the average energy loss per ICS interaction. The dataset consists of 4000 electrons with an initial energy of 10~TeV. The significant energy loss per ICS interaction leads to a large dispersion in the final electron energy, preventing the formation of a spectral peak.}
\label{fig: final electron energy vs average loss per ICS interaction}
\end{figure}

To produce an accurate model, we inject $\sim$30,000 electrons with an initial energy distribution following Eqn.~\ref{eq:pulsarspectrum} and time-evolution from Eqn.~\ref{eq:luminosityvstime}. We include electrons from 100~GeV to 1000~TeV in 5000 logarithmic bins. We bin the final electron energies into 30 logarithmic bins between 100~GeV and 10~TeV. We generate several alternative datasets for parameter space tests described in Appendices~\ref{app: further analysis} and \ref{app: isotropic}.

We compare our stochastic method to an analytic calculation that produces a sharp spectral cutoff. We use Eqn.~\ref{eq: analytic flux}, which gives the differential flux for electrons injected a time \mbox{$t' = 342$~kyr$-t$} ago for a pulsar at distance $d$ from Earth with an injection spectrum following Equation~\ref{eq:pulsarspectrum} with 320 logarithmic bins between $100$~GeV and $1000$~TeV. To model electrons injection continuously over time, we sum the results for single $\delta$-function injections for $t'=0$ -- $342$~kyr and normalize the flux in each time step following Eqn.~\ref{eq:luminosityvstime}.

\begin{figure}
\centering
\includegraphics[width=0.49\textwidth]{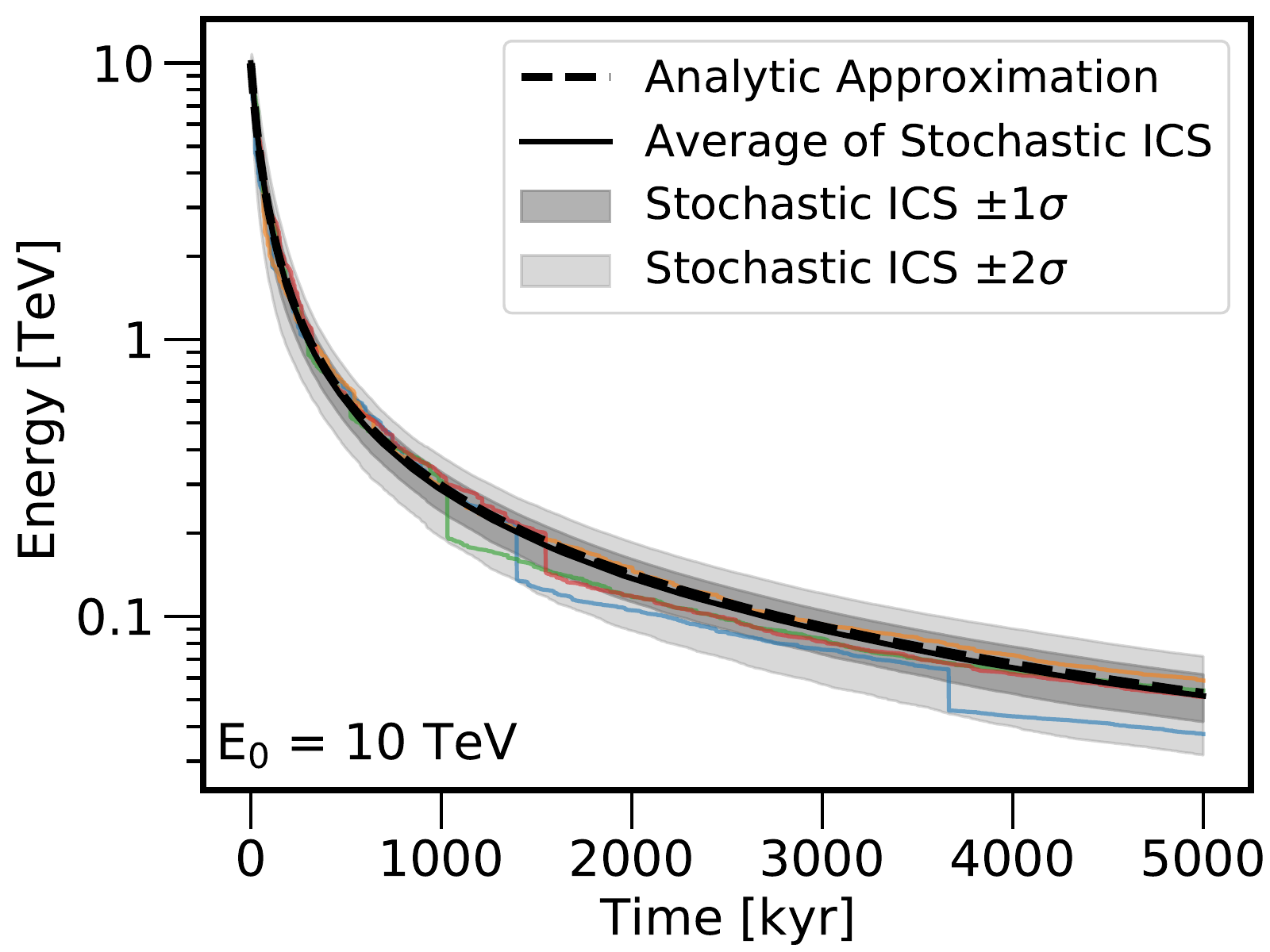}
\caption{The energy of an electron with an initial energy of 10~TeV over the 5~Myr duration of our simulation. The average (black solid line), and 1$\sigma$ (dark gray) and 2$\sigma$ (light gray) deviations are shown and compared with the electron energy from the analytic approximation. We show colored curves depicting a few typical events.
}
\vspace{-0.4cm}
\label{fig: electron energy vs time}
\end{figure}

Only a small fraction of electrons produced by the pulsar reach Earth. Our stochastic model does not include diffusion and provides the total electron power from the pulsar. To compare this with the analytic calculation, which gives the electron flux at radius $r$, we integrate the analytic flux over all space. In order to compare our results to an observed electron spectrum, one would need to take diffusion into account to obtain the spectrum at Earth. We stress that this approach does not affect our qualitative results, as the spectral feature depends on cooling, not diffusion. We discuss diffusion further in Appendix~\ref{app: diffusion}.

\section{Results}
Figure~\ref{fig: final electron energy vs average loss per ICS interaction} shows the fundamental process that disperses the final electron energy. Starting with electrons at an initial energy of 10~TeV, we show the average energy lost per ICS event and the final energy at 342~kyr. On average, each ICS event removes $\sim$55~GeV from an electron, indicating that stochastic variations in the number and strength of ICS events lead to detectable dispersion in the electron energy. For these initial conditions, the dispersion is 872 $\pm$ 145~GeV, where we note that 872~GeV represents the average energy of the electron population, and 145~GeV represents the dispersion in the energies of single electrons around this average.

Figure~\ref{fig: electron energy vs time} shows the time-evolution of the electron energy from Figure~\ref{fig: final electron energy vs average loss per ICS interaction}. The initial dispersion is small because most electrons have not yet had an ICS event. However, the dispersion increases quickly. At 5~Myr, the electron energies are spread between 42 -- 62~GeV at $1\sigma$ and 32 -- 72~GeV at $2\sigma$. We note that the average final energy in our stochastic model is 52~GeV, which is essentially equivalent to the 53~GeV final energy in the analytic case.

Combining these features, we show our main result in Figure~\ref{fig:1}. The standard analytic approximation has three features: (1) a sharply rising electron spectrum at low energies, where the electrons are not cooled and maintain their injection spectrum, (2) a steep drop at a critical energy that corresponds to the efficient cooling of higher-energy electrons produced near $t=0$, and (3) a softer high-energy spectrum produced by cooled electrons emitted at later times.

Our exact solution includes the first and third features, because the \emph{average} ICS cooling is correct in the analytic approximation. However, the sharp spectral feature is smoothed out by the different energy losses experienced by individual electrons. Notably, this effect is much larger than the energy resolution of current cosmic-ray experiments such as AMS-02, CALET and DAMPE~\cite{PhysRevLett.113.121101, DAMPE:2017fbg, Adriani:2018ktz}. \\

\subsection{Discussion} We have shown that the standard analytic approximation for ICS cooling (Eqn.~\ref{eq:standardenergylosses}) induces an erroneous spectral feature in the local electron and positron fluxes. A proper treatment accounting for the stochasticity of ICS does not produce this feature. Physically, this stems from the fact that electrons only interact with a small random sample of the photon field, cooling to an energy that is described by a probability distribution function rather than an exact value. 

We stress that while this stochastic effect is most pronounced at high energies, it is not physically related to the kinematic effects of Klein-Nishina suppression, but is purely due to the statistics of ICS interactions. This fact is clearest in Figure~\ref{fig: electron energy vs time}, where we see that the significant dispersion in the final electron energies continues to nearly $\sim$50~GeV at 5~Myr, far lower than the standard Klein-Nishina range.

Our results are applicable to more diverse phenomena than the positron excess. ICS cooling cannot produce spectral features, owing to its inherent stochasticity. Our results hold for any system where particles are stochastically cooled, including \emph{e.g.}, supernova models of the electron and positron fluxes~\cite{Huang:2017egk}. Similar effects stemming from catastrophic energy loss processes have been discussed in the case of p-$\gamma$ interactions~\cite{1990Ap&SS.167...93A} and secondary antiproton production~\cite{Moskalenko:2001ya}.

Interestingly, a peaked local electron spectrum is possible if cooling were dominated by synchrotron, rather than ICS -- making a spectral peak a diagnostic for the energy loss process. However, for local studies, this is an academic concern. Any source close enough to contribute to the electron flux that had sufficient synchrotron cooling to dominate ICS losses would have already been detected in radio data. \\

\subsection{Effect on Pulsar Models} Our results have significant implications for pulsar models of the positron excess. Early studies realized that the number of pulsars that contribute to the excess is energy-dependent~\cite{Hooper:2008kg}, due to the fact that the energy-dependence of diffusion ($D$~$\propto$~$E^{\delta}$ with \mbox{$\delta \sim 0.4$~\cite{Korsmeier:2021bkw}}) is weaker than the $E^{-1}$ energy-loss timescale for synchrotron and ICS. This means that low-energy electrons can travel farther from pulsars before cooling. This was examined quantitatively in Ref.~\cite{Bitter:2022uqj, Cholis:2021kqk}, and is not affected by our result.

Several recent studies have produced detailed models of Milky Way pulsars to determine the characteristics of systems that contribute to the positron flux~\cite{Cholis:2017ccs, Cholis:2018izy, Evoli:2020szd, Orusa:2021tts, Cholis:2021kqk}. Because these models produce $\chi^2$ fits to AMS-02 (and also DAMPE and CALET) data over a large energy range, they quantitatively constrain spectral features from individual pulsars. Because the observed electron and positron spectra are smooth, these studies tend to rule out models where a few young pulsars would produce large spectral features. 

We note that each of these models have many free parameters, and treat systematic errors differently. Thus, it is difficult to determine how the ICS approximation affects each. However, any study that uses an analytic ICS model will produce artificially strong constraints on the contribution from nearby pulsars with ages between $\sim$100--1000~kyr, because such systems would produce spectral features between 100--1000~GeV that have not been observed. This constraint is weaker at lower energies because a larger number of pulsars contributes to the excess, and weaker at higher energies because of the larger uncertainties in cosmic-ray data.

These constraints are in modest tension with TeV halo data, which show that Geminga and Monogem (among others) are powerful electron accelerators. Studies have discussed several effects that could be invoked in standard ICS models to decrease their spectral bumps, including: (1) inhomogeneities in energy-losses~\cite{Malyshev:2009tw} or diffusion~\cite{Do:2020xli} that may affect the uniformity of the electron flux reaching Earth, (2) the effective trapping and cooling of young electrons within the pulsar wind nebulae~\cite{2018NPPP..297..106G, Evoli:2020szd}, (3) changes to the pulsar spin-down timescale which increase the fraction of electrons that are accelerated at late times~\cite{Cholis:2021kqk}, or (4) the energetic dominance by an extremely young pulsar with a spectral bump above the energy of current data~\cite{Orusa:2021tts}. Our models do not necessarily reject such ideas but they do diminish the need for such models. However, it is possible that the spectral feature would be even more smoothed out by these possible effects.

To be clear, our results show that pulsars \emph{do not} produce sharp spectral features -- a result which is based only on known particle physics. Our results re-open the possibility that only a small number of pulsars produce the positron excess at high-energies. Additionally, our analysis indicates that current (or even future~\cite{Schael:2019lvx}) studies of the electron and positron fluxes will not find sharp spectral features that can be used to constrain the age or proximity of nearby sources.

\subsection{Effect on Dark Matter Searches} Dark matter particles that annihilate into $e^+e^-$ pairs or other leptonic states are predicted to produce features in the cosmic-ray positron spectrum. This dark matter contribution would be subdominant, rather than accounting for the majority of the positron excess, and includes a sharp cutoff corresponding to the mass of the dark matter particle~\cite{Turner:1989kg}. This spectral cutoff is intrinsic to the electron production process (and not caused by cooling). Our results do not affect this conclusion. 

Excitingly, our analysis indicates that there is no standard astrophysical mechanism capable of producing a sharp feature in the local electron spectrum. The detection of such a feature, in this case, would serve as incontrovertible evidence of dark matter annihilation, or another novel physics process.

\subsection{Diffusion} For clarity, this paper focuses on cooling and ignores diffusion. Of course, to compare our results with an observed positron flux taken at the specific solar position, one would need to directly model the diffusion of cosmic-rays from Geminga to Earth. However, we note that diffusion cannot ``re-create" a spectral peak for two reasons: (1) the energy-dependence of diffusion is monotonic. To create a feature, the diffusion coefficient would need a sharp ``spike" at a specific energy, (2) any such spike would affect all cosmic rays at a given rigidity, producing a sharp feature in all cosmic ray data that is ruled out. We provide more details in Appendix~\ref{app: diffusion}.

\vspace{-0.3cm}
\section*{Acknowledgements}
\noindent We thank Felix Aharonian, John Beacom, Ilias Cholis, Pedro De la Torre Luque, Carmelo Evoli, Ottavio Fornieri, Dan Hooper, Dmitry Khangulyan, Michael Korsmeier, Silvia Manconi, Igor Moskalenko, Payel Mukhopadhyay, Alberto Oliva, Luca Orusa, Stefano Profumo, Stefan Schael, Pasquale Serpico and Takahiro Sudoh for helpful comments. TL is supported by the Swedish National Space Agency under contract 117/19, the Swedish Research Council under contracts 2019-05135 and 2022-04283 and the European Research Council under grant 742104. This project used computing resources from the Swedish National Infrastructure for Computing (SNIC) under project Nos. 2021/3-42, 2021/6-326 and 2021-1-24 partially funded by the Swedish Research Council through grant no. 2018-05973.

\appendix

\section{Cosmic-Ray Electron Acceleration and Propagation}\label{app: standardmodel}
The most common treatment of electron acceleration, propagation, and cooling in pulsars is as follows. The pulsar is born at time $t$=0 (matching the supernova) and immediately begins to inject e$^+$e$^-$ pairs with a flux that is proportional to its spin-down power. The spin-down power is calculated using a simple model where the pulsar is treated as a mis-aligned rotating dipole, with a luminosity:

\begin{equation}
    \label{app:eq:luminosity}
    L(t) = \eta L_0 \left ( 1+\frac{t}{\tau} \right )^{-(n+1)/(n-1)}
\end{equation}

\noindent where L$_0$ is the power at time $t$=0 and is normalized to the pulsar kinetic energy, $n$ is the breaking index, which is usually set to 3, $\eta$ is an efficiency parameter, which is typically assumed to be constant with a typical range of 0.01--1 (though see~\cite{2018NPPP..297..106G, Evoli:2020szd}), and $\tau$ sets the energy loss timescale, which can be calculated in the dipole model as:

\begin{equation}
    \tau = \frac{3c^3 I P_0^2}{4 \pi^2 B^2 R^6}
\end{equation}

\noindent The specific value for $\tau$ can change significantly depending on the initial period and magnetic field strength of individual pulsars~\cite{Sudoh:2019lav, Fornieri:2019ddi, Orusa:2021tts}. Here, we adopt $\tau$~=~9.1~kyr, as calculated by Ref.~\cite{Hooper:2017gtd} for the Geminga pulsar. Because this time scale is short compared to the $\sim$100~kyr age of pulsars that contribute to the positron excess, some studies (e.g. Refs~\cite{Profumo:2008ms, Cholis:2021kqk}), further simplify the modeling by assuming that pulsars instantaneously inject all their energy at time $t$~=~0.

The pulsar spectrum is typically modeled as a power-law with an exponential cutoff. However, the actual mechanism producing this acceleration is unclear, as there are two possibilities. The first is direct e$^\pm$ pair production and acceleration at the pulsar magnetosphere, a process which may be either efficient or inefficient depending on the pulsar magnetosphere model, and which can continue to energies well-above 1~TeV~\cite{Philippov:2017ikm, Kalapotharakos:2017bpx}. The second option is that the electrons are originally produced in the pulsar-magnetosphere, but are then re-accelerated (and their spectrum is reset) as they transit through the termination shock of the surrounding pulsar wind nebula~\cite{Gaensler:2006ua, 2011ApJ...726...75S, Cerutti:2020qav}. Future observations of systems that do not include pulsar wind nebulae (e.g., milisecond-pulsars) could potentially distinguish these possibilities~\cite{Hooper:2018fih, Hooper:2021kyp}.

In either case, the electron pairs diffuse away from the pulsar/pulsar wind nebula, following a process that is typically treated using the diffusion/convection equation of the form:

\begin{multline}
    \frac{\partial}{\partial t}\frac{d\psi}{dE} \left(E,r,t\right) = \vec{\nabla} \cdot \left[ D(E) \nabla \frac{d\psi}{dE} - v_c \frac{d\psi}{dE} \right] \\ + \frac{d}{dE}\left [\frac{dE}{dt}\frac{d\psi}{dE} \right ] + \delta(r)Q_0(r,t,E)
\end{multline}

\noindent where $D(E)$ is the diffusion coefficient, which is typically normalized to fit cosmic-ray secondary-to-primary ratios, with a typical energy dependence $D(E)$ $\propto$ $E^{\alpha}$ with $\alpha$ in the range 0.33 -- 0.5~\cite{Korsmeier:2021bkw}, $v_c$ is a convection term which we will set to 0 in this study, the energy derivative accounts for energy losses due to synchrotron and ICS, and finally $Q_0$ is the source term, which is a spectrum-dependent normalization constant that is set such that the integral of the pulsar emission matches the pulsar luminosity from Eq.~\ref{app:eq:luminosity}.

While this transport equation must typically be solved numerically, there is an commonly-employed analytic formula in the case that the cosmic-ray injection rate is a delta-function in time, which is given by:

\begin{equation}
\label{eq: analytic flux}
    \frac{dN}{dE} = \frac{Q(E)}{\pi^{3/2} d^3}(1-bt'E)^{(\alpha-2)}\left ( \frac{d}{D_{{\rm diff}}} \right )^3 e^{-(d/D_{{\rm diff}})^2}
\end{equation}
where $d$ is the distance from the pulsar to Earth, $t'$ is the time since electron injection (e.g., 342~kyr - $t$) and $D_{\rm diff}$ is:

\begin{equation}\label{eq: diffusion distance}
D_\text{diff}\left(E_e, t'\right) \approx 2\sqrt{D\left(E_e\right)t'\frac{1 - \left(1-E_e/E_\text{max}\right)^{1-\delta}}{\left(1-\delta\right)E_e/E_\text{max}}}.
\end{equation}

\noindent $E_\text{max}$ is the maximum energy that can be lost due to conversation of energy, given by $E_\text{max} \approx 1/\left(bt'\right)$, and $D\left(E_e)\right)$ is the diffusion coefficient for a specific electron energy given by:

\begin{equation}\label{eq: diffusion}
D\left(E_e\right)= D_0 \times \left(\frac{E_e}{1\text{ GeV}}\right)^\delta,
\end{equation}
where we take $D_0 = 2\times10^{28}$ cm$^2$/s at an electron energy of 1 GeV and a diffusion spectral index of $\delta=0.4$~\cite{Hooper:2017gtd}. 

Following Equation~\ref{eq:standardenergylosses}, the energy loss rate is typically written as a continuous process, which is given by:

\begin{equation}\label{eq: energy loss rate}
\frac{dE}{dt} = - b\left(E\right) \times \left(\frac{E}{1\text{ GeV}}\right)^2,
\end{equation}

where $b$ is typically calculated as:

\begin{multline}\label{eq: b}
b\left(E\right) = 1.02\times10^{-16}\times \\
\left(\sum_i\frac{\rho_i}{\text{eV/cm}^3}\,S_i\left(E\right) + \rho_\text{mag}\left(\frac{B}{3\,\mu\text{G}}\right)^2 \right) \text{ GeV/s}.
\end{multline}

\noindent though we note that there are more accurate analytic prescriptions in the literature which take into account the energy-dependence of $b$~\cite{Khangulyan:2013hwa}. In particular, this cross-section is inhibited at high-energies due to Klein-Nishina suppression, which decreases the incidence angles over which a photon and electron have a high-interaction probability. Studies have either used an exact calculation of the Klein-Nishina suppression~\cite{Blumenthal:1970gc}, or utilized a simplified suppression factor $S(E)$ first calculated in~\cite{2010NJPh...12c3044S}. Throughout this paper, we utilize the exact solution for the Klein-Nishina suppression calculated at each $\gamma$-ray and initial photon energy, following Ref.~\cite{Blumenthal:1970gc} and given by:

\begin{equation}
\frac{dE}{dt} = \frac{12c\sigma_T E^2}{m_e c^2} \int_0^\infty J\left(\Gamma\right) \nu n\left(\nu\right) d\nu, 
\end{equation}
where $\sigma_T$ is the Thomson cross section, $\nu$ is the energy of the ISRF photons and $n\left(\nu\right)$ represents their energy spectrum (in 1/(eV cm$^3$), $\Gamma = 4\nu\gamma/\left(m_e c^2\right)$ and \mbox{$q = \nu_s/\left(\Gamma\left(\gamma m_e c^2 - \nu_s\right)\right)$} takes into account the final $\gamma$-ray energy $\nu_s$, and $J\left(\Gamma\right)$ is in integral given by

\begin{equation}
J\left(\Gamma\right) = \int_0^1 \frac{q G\left(q, \Gamma\right)}{\left(1+\Gamma q\right)^3} dq.
\end{equation}

While we utilize the exact solution for the Klein-Nishina cross-section, we stress that the difference between the approximate and exact solutions for Klein-Nishina suppression is not relevant for our study -- because both are calculated within the continuous energy-loss formalism. Any method utilizing Equation~\ref{eq:standardenergylosses} will incorrectly induce an electron spectral peak regardless of whether an exact or approximate model for the Klein-Nishina effect is employed. 

The effect of ICS cooling depends sensitively on the model for the ISRF. There are many possibilities, including full spectral models based on multiwavelength observations as well as approximate models that bin the ISRF into a few major components with specified energies. In this study, we take the temperatures \mbox{$T_\text{UV} = 20\times10^3$~K}, \mbox{$T_\text{optical} = 5\times10^3$~K}, \mbox{$T_\text{IR} = 20$~K} and \mbox{$T_\text{CMB} = 2.7$~K}, and the energy densities \mbox{$\rho_\text{UV} = 0.1$~eV/cm$^3$}, \mbox{$\rho_\text{optical} = 0.6$~eV/cm$^3$}, \mbox{$\rho_\text{IR} = 0.6$~eV/cm$^3$}, \mbox{$\rho_\text{CMB} = 0.26$~eV/cm$^3$}, \mbox{$\rho_\text{mag} = 0.224$~eV/cm$^3$} (corresponding to a magnetic field strength of $B = 3\,\mu$G).~\cite{Hooper:2017gtd}. We again stress that differences between these approaches are not responsible for the feature we identify in the main text.
\color{black}

\begin{figure}[tbp]
\centering
\includegraphics[width=0.48\textwidth]{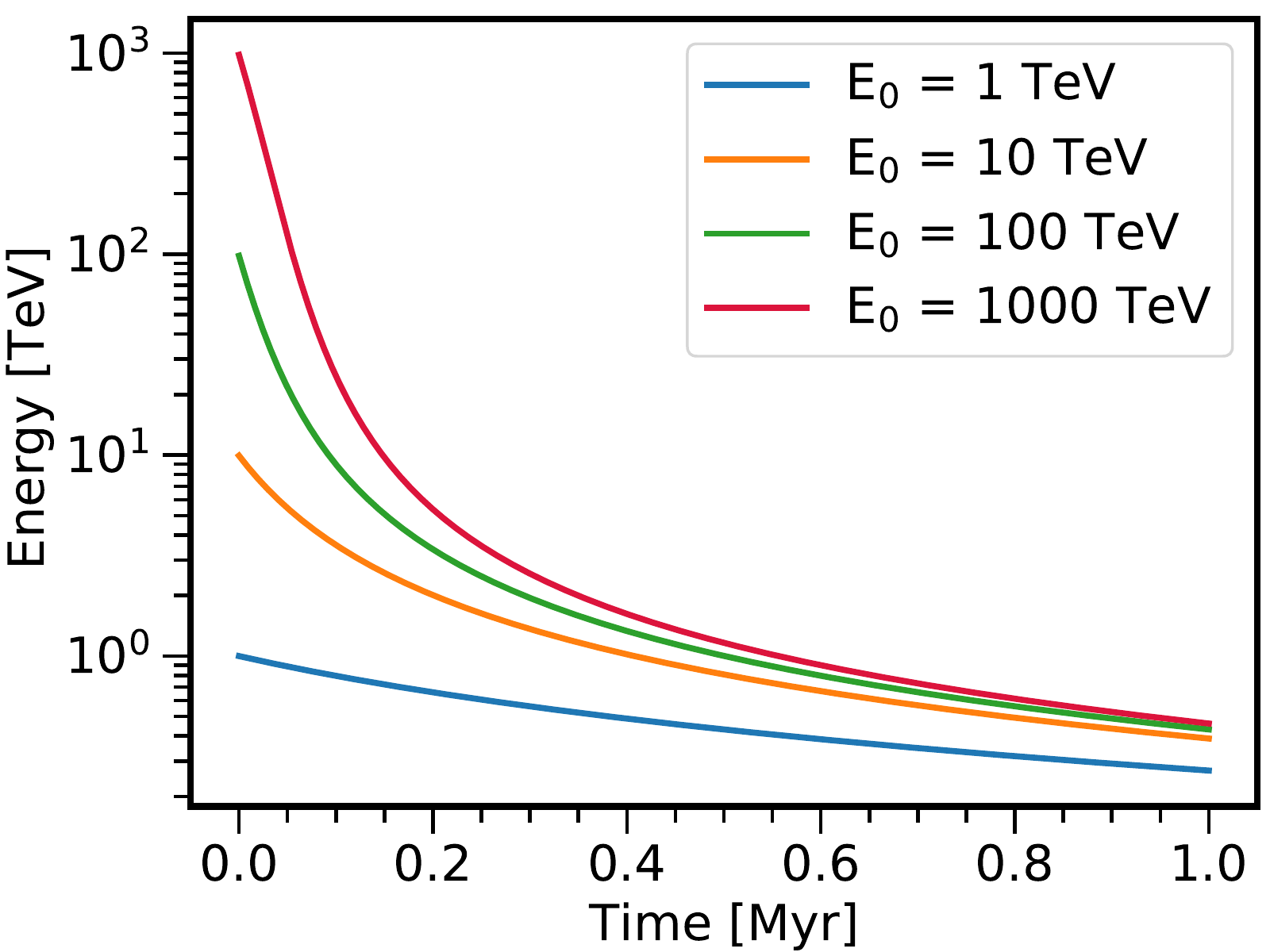}
\caption{The evolution of the electron energy over 1~Myr in the standard analytic approximation for electrons with different initial energies (1~TeV (blue), 10~TeV (orange), 100~TeV (green) and 1000~TeV (red)). Electrons with higher initial energies cool more rapidly than lower energy electrons, causing the electrons to cool down to similar energies over time, producing a spectral feature.}
\label{fig: energy loss vs time}
\end{figure}

Figure~\ref{fig: energy loss vs time} shows the energy losses for the analytic approximation over 1~Myr for electrons with different initial energies. Electrons with high energies cool faster than electrons with low energies, and over time cool down to similar energies. After 1~Myr, the electrons with initial energies between 10 and 1000~TeV have roughly cooled to the same energy, 0.42~TeV. The fact that a larger fraction of initial lines converge at later times drives the fact that models of pulsar contributions provide increasingly peaky spectral features for older pulsars. \\

\section{Further Analysis of Stochastic Inverse-Compton Scattering} \label{app: further analysis}
\begin{figure}[h]
\begin{minipage}[t]{0.48\textwidth}
\centering
\includegraphics[width=0.78\textwidth]{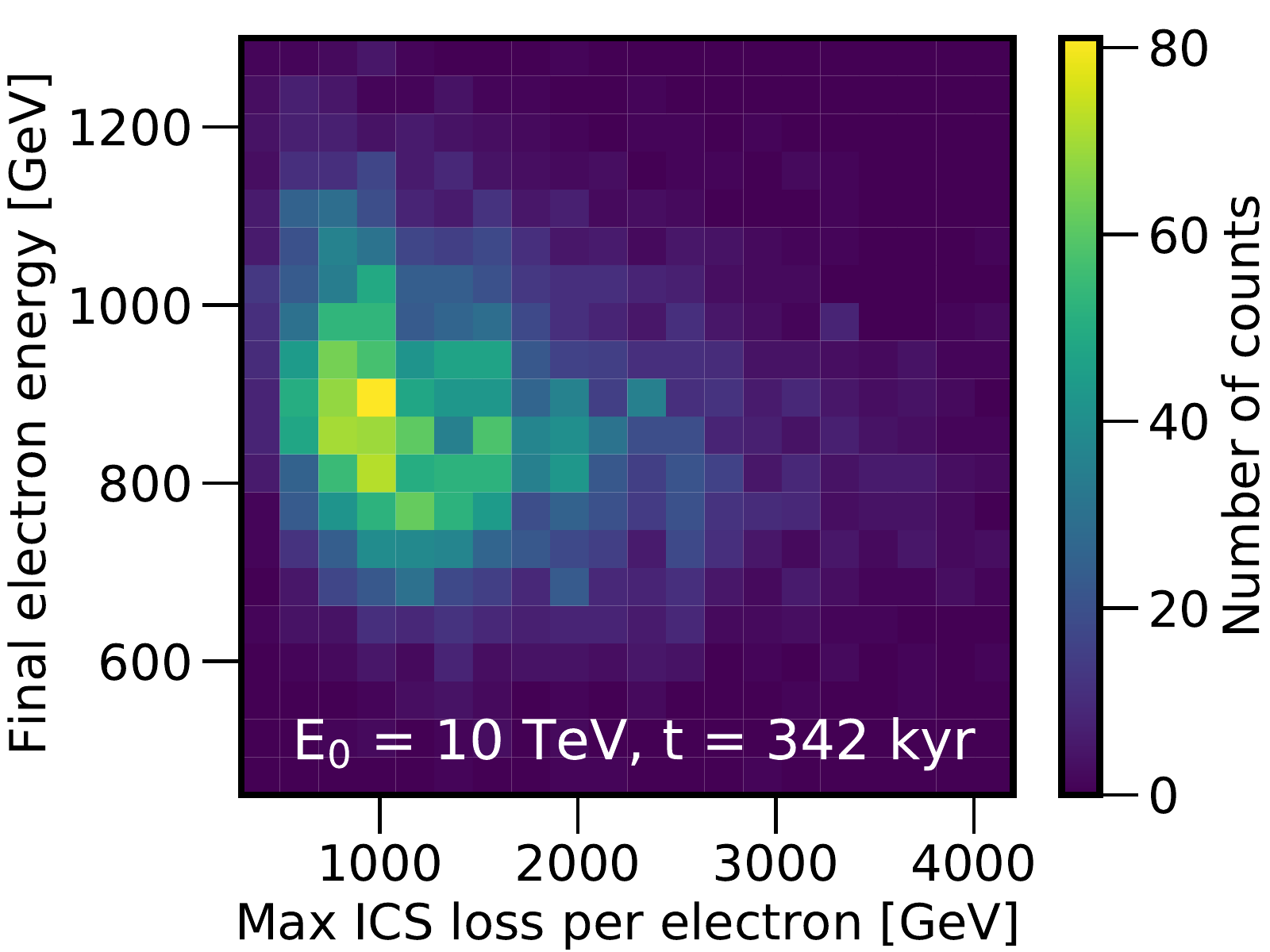}
\caption{The final electron energy compared to the maximum energy loss for an electron with an initial energy of 10~TeV, cooled over 342~kyr. The vast majority of electrons have at least one ICS interaction which removes over 1~TeV from the electron.}
\label{fig: final electron energy vs max ICS loss}
\vspace{0.1cm}
\end{minipage}
\hfill
\begin{minipage}[t]{0.48\textwidth}
\centering
\includegraphics[width=0.78\textwidth]{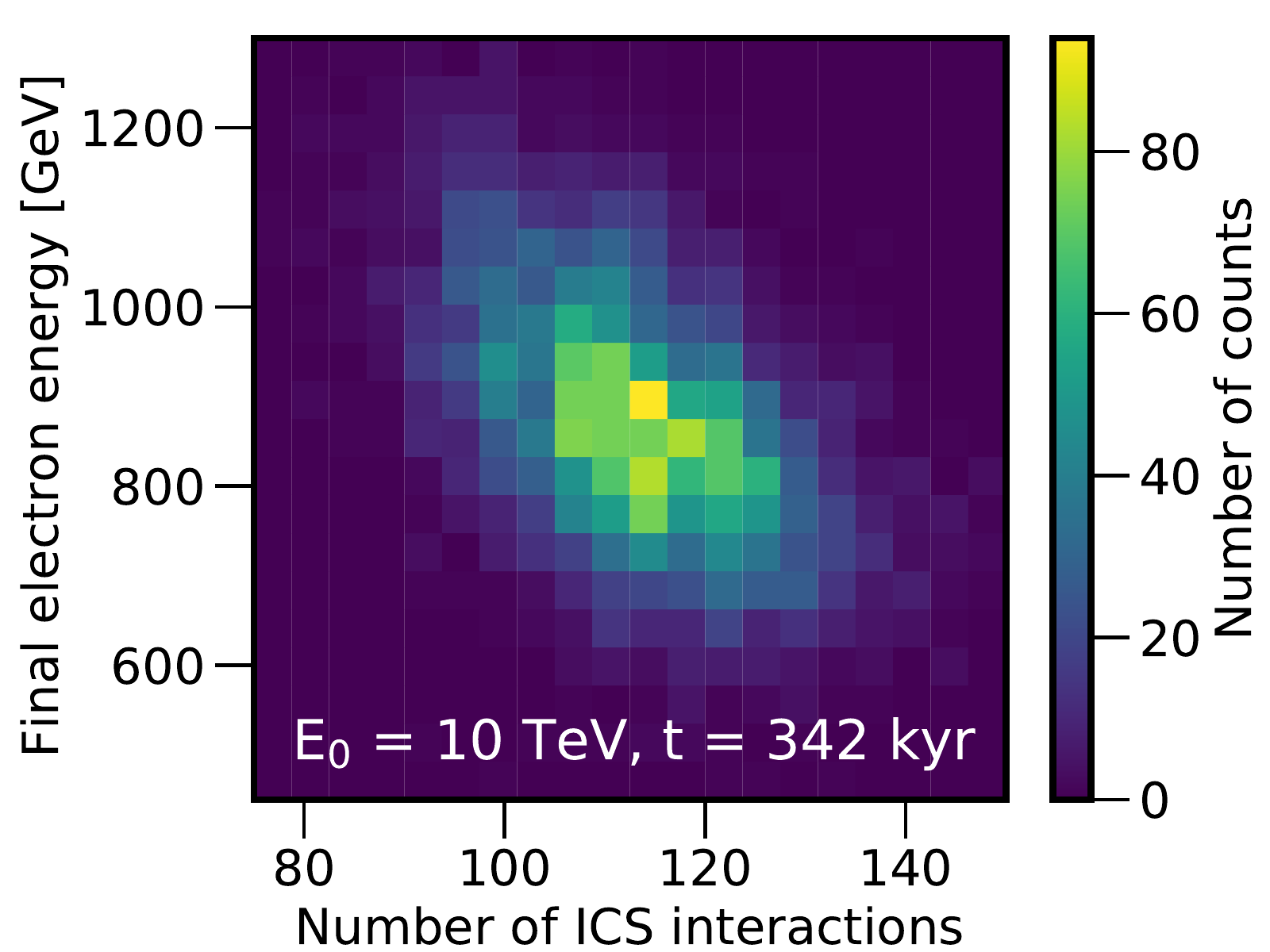}
\caption{The final electron energy compared to the number of ICS interactions an electron with initial energy 10~TeV experiences over 342~kyr.}
\label{fig: final electron energy vs number of ICS interactions}
\end{minipage}
\hfill
\begin{minipage}[t]{0.48\textwidth}
\centering
\includegraphics[width=0.78\textwidth]{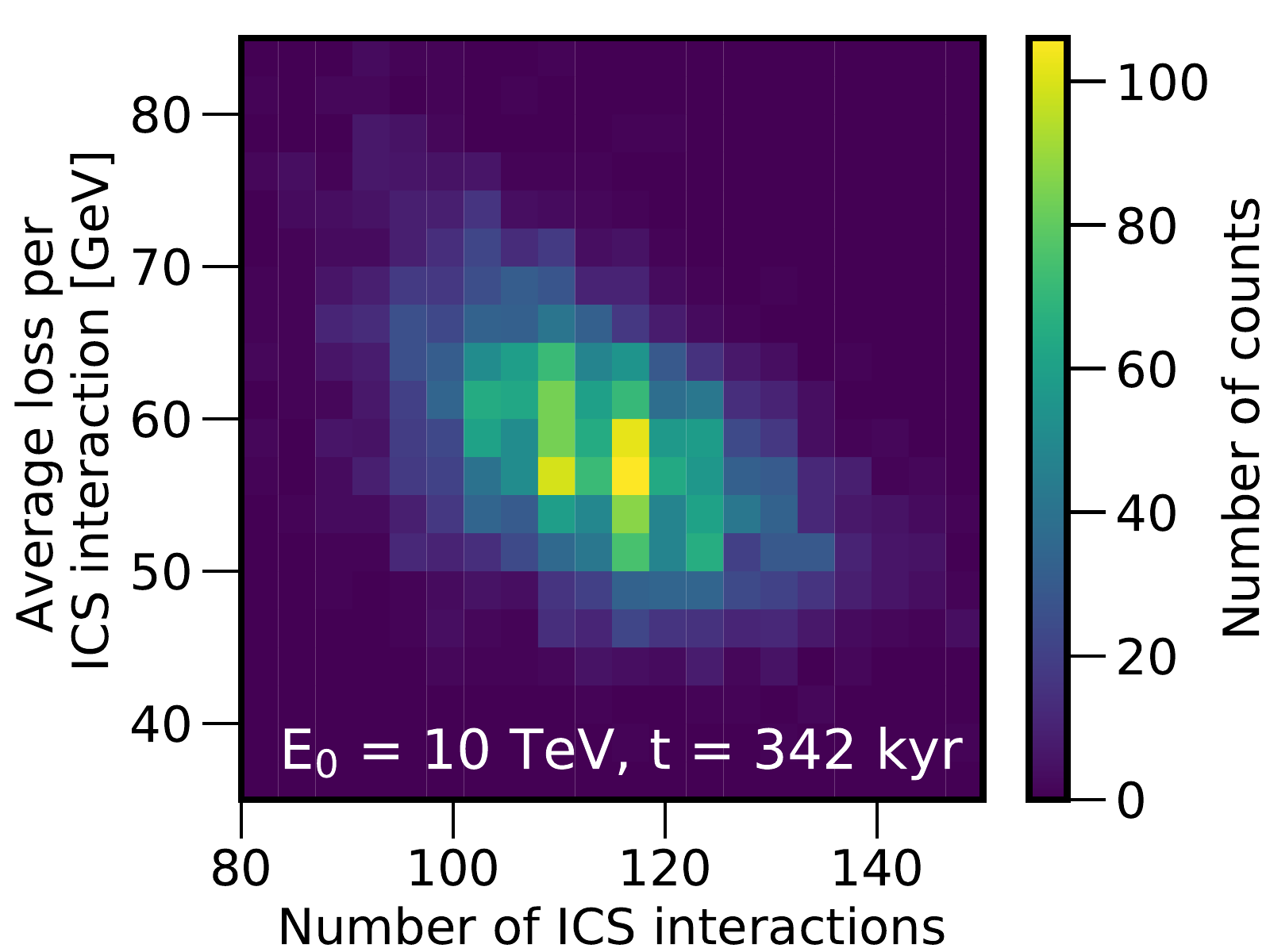}
\caption{The average energy loss per ICS interaction for an electron compared to the number of ICS interactions an electron with initial energy 10~TeV experiences over 342~kyr.}
\label{fig: average electron energy vs number of ICS interactions}
\end{minipage}
\end{figure}

Figures~\ref{fig: final electron energy vs max ICS loss}--\ref{fig: average electron energy vs number of ICS interactions} show several additional plots depicting the distribution of interactions that our electron population undergoes while interacting with the ISRF. These results correspond to the simulations produced in Figures~\ref{fig: final electron energy vs average loss per ICS interaction} and~\ref{fig: electron energy vs time} of the main text, meaning that they simulate 1000 electrons with an initial energy of 10~TeV and produce results for a simulation that lasts 342~kyr, corresponding to the age of Geminga. 

In Figure~\ref{fig: final electron energy vs max ICS loss}, the final electron energy is shown compared to the maximum energy loss an electron has experienced in an ICS interaction. The maximum energy loss for each individual electron lies between approximately 600 to 1700~GeV, which is a significant fraction of the total electron energy. Over 342~kyr, these electrons undergo relatively few ICS interactions, typically 90 -- 130 interactions, as shown in Figure~\ref{fig: final electron energy vs number of ICS interactions} for the final electron energy against the number of ICS interactions per electron. The large spread in the energy loss per ICS interactions and the large variation in the number of interactions results in the large spread of final electron energies. Finally, Figure~\ref{fig: average electron energy vs number of ICS interactions} shows the average energy loss per ICS interaction compared to the number of ICS interactions it undergoes.

\begin{figure}[tbp]
\begin{minipage}[t]{0.48\textwidth}
\centering
\includegraphics[width=0.9\textwidth]{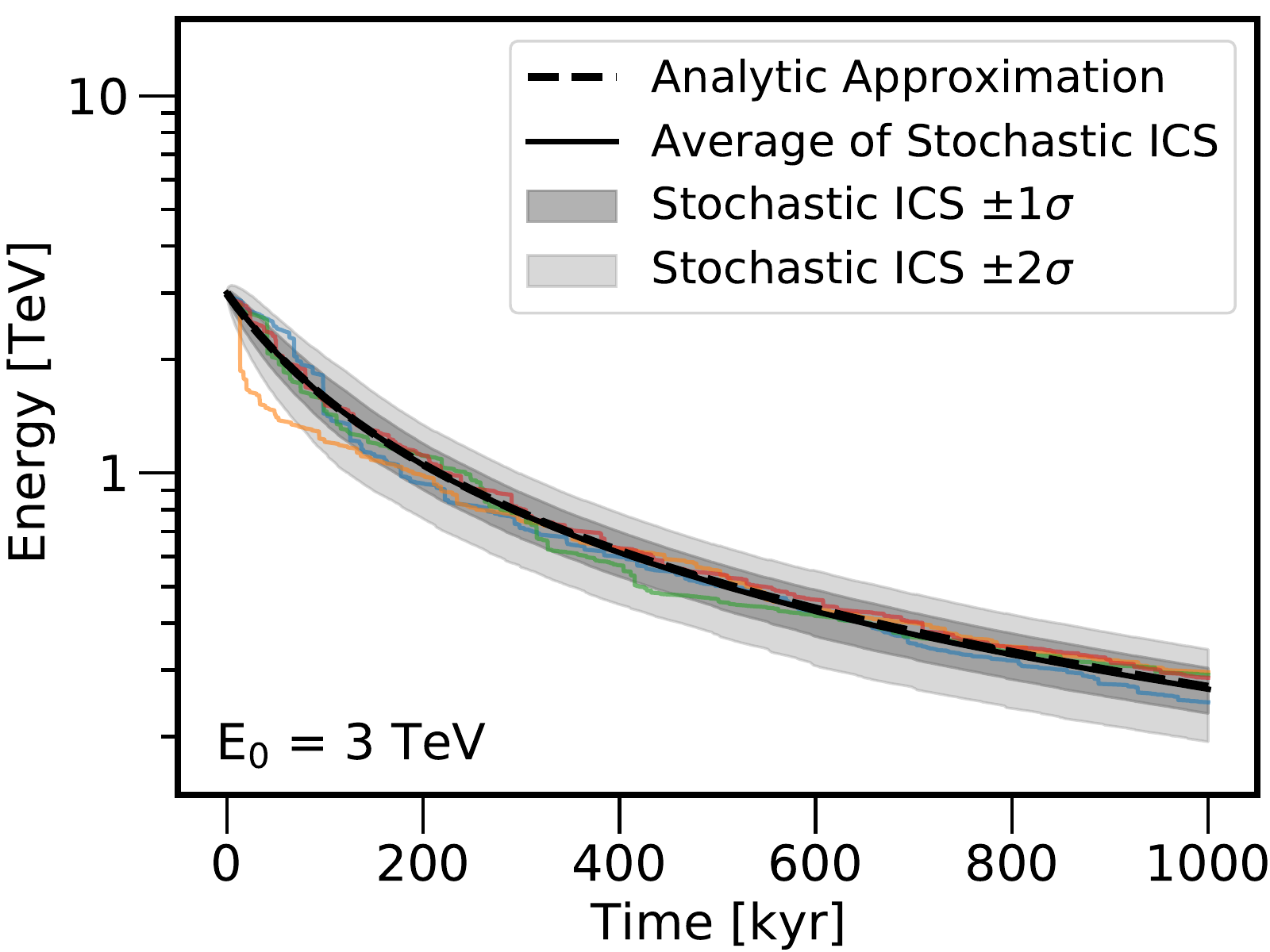}
\caption{The electron energy as a function of time for initial electron energy of 3 TeV. A similar dispersion is seen despite the lower starting energy, which demonstrates that our effect is not dependent on interactions in the strong Klein-Nishina limit.}
\label{fig: electron energy vs time 3 TeV}
\end{minipage}
\hfill
\begin{minipage}[t]{0.48\textwidth}
\centering
\includegraphics[width=0.9\textwidth]{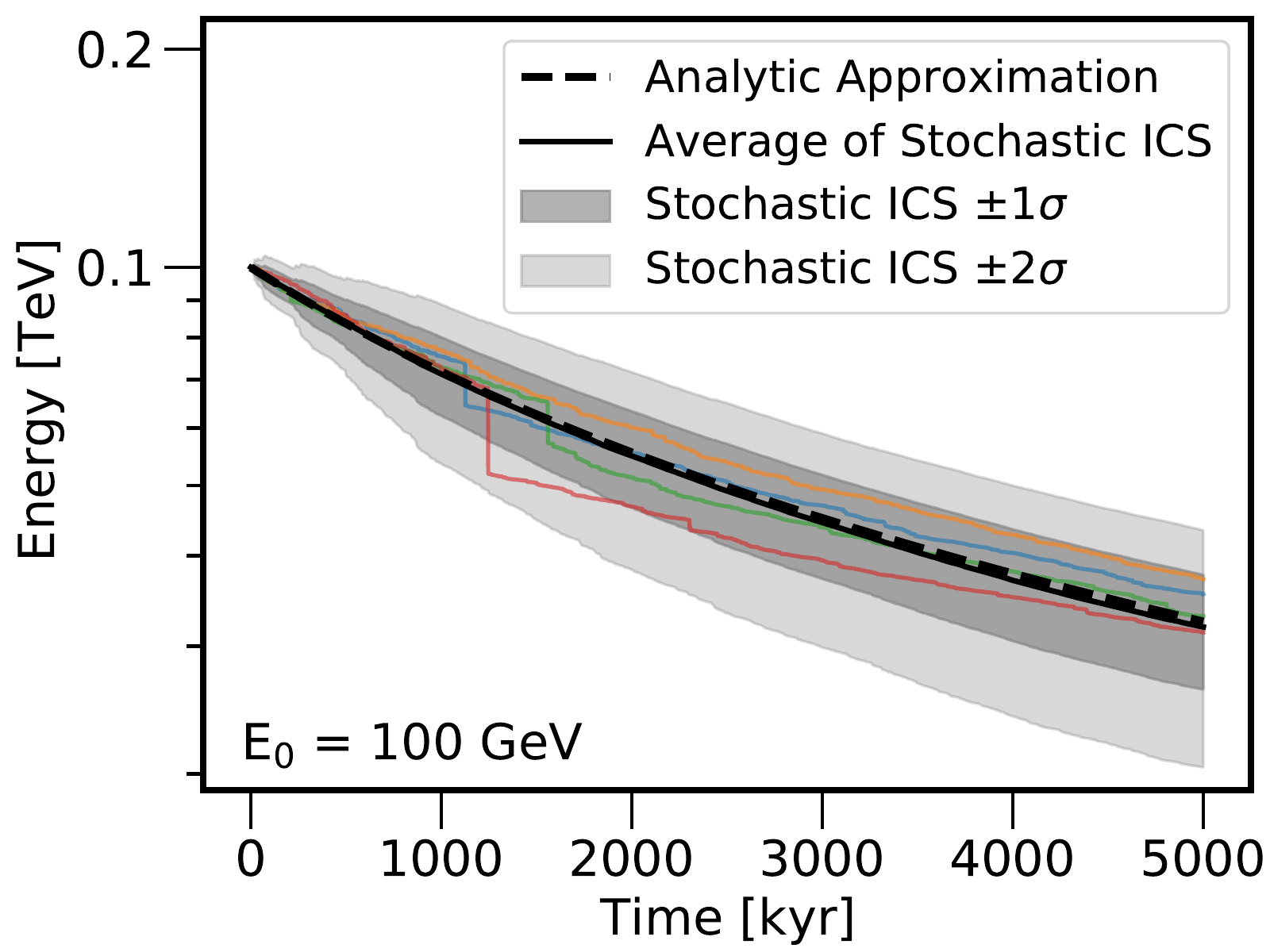}
\caption{The energy evolution over 5~Myr for electrons with an initial energy of 100~GeV. We note that there is a significant energy dispersion in the final energy, which exceeds 15\% at 1$\sigma$. Because the ICS interactions of 100~GeV electrons fall primarily within the Thomson regime, this analysis illustrates that Klein-Nishina effects are not responsible for the stochastic energy loss effect that we demonstrate.}
\label{fig: 5Myr 100 GeV}
\end{minipage}
\hfill
\begin{minipage}[t]{0.48\textwidth}
\centering
\includegraphics[width=0.9\textwidth]{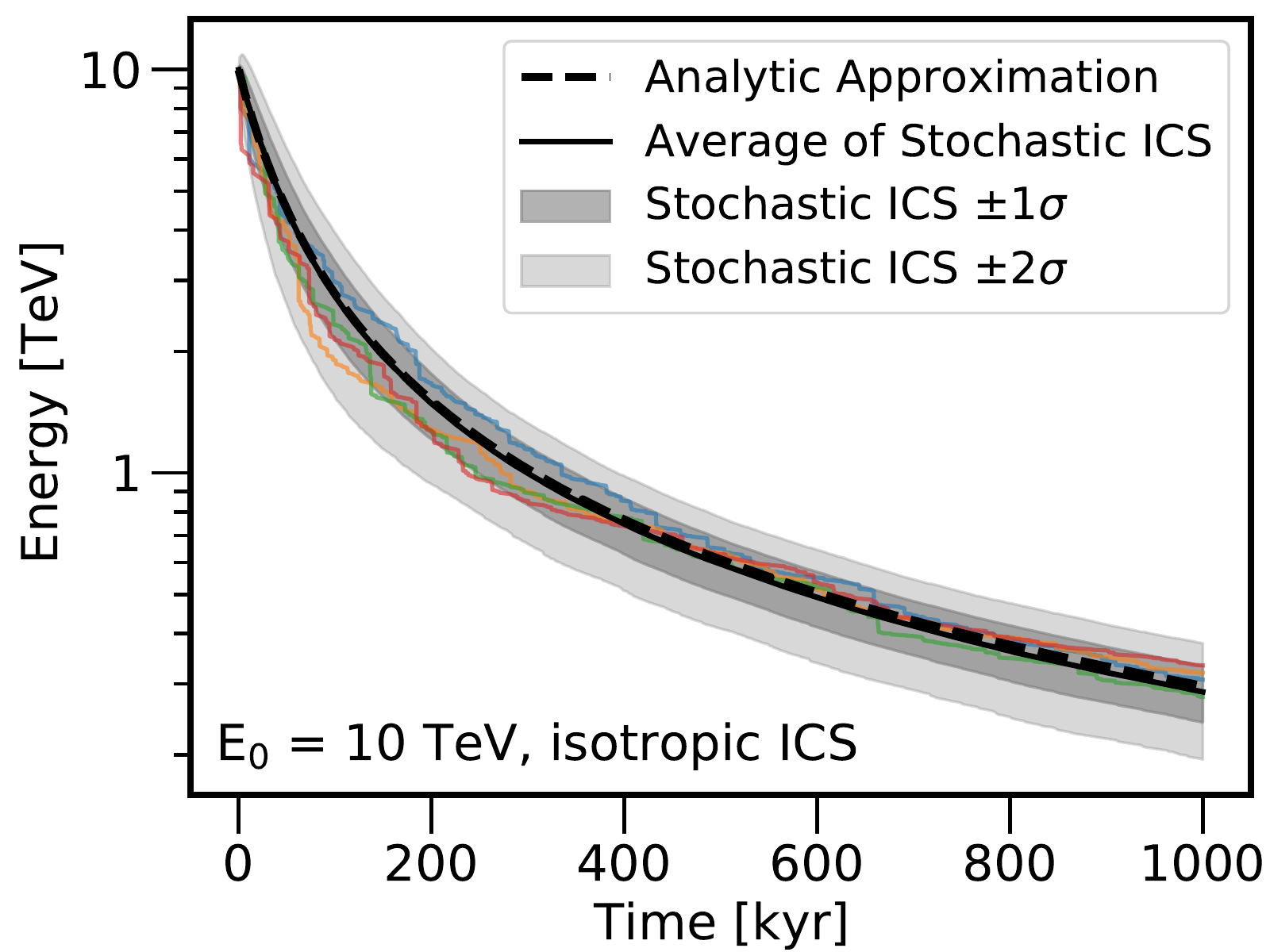}
\caption{The electron energy against time for initial electron energy of 10 TeV for the isotropic ICS calculation. Using an isotropic ICS model has almost no effect on the cooling rates, or energy dispersion of our electron population, which is expected because both the ISRF and electron flux are isotropic.}
\label{fig: electron energy vs time isotropic}
\end{minipage}
\end{figure}

Figure~\ref{fig: electron energy vs time 3 TeV} shows the evolution of the electron energy over 1000~kyr, similar to Figure~\ref{fig: electron energy vs time}, but with an initial electron energy of 3~TeV for 1000 electrons. The final average energy of the exact stochastic ICS calculation is 267~GeV (black-solid) with an energy spread of 230--304~GeV at $1\sigma$ and 194--340~GeV at $2\sigma$. The final energy in the standard analytic approximation is 271~GeV (black-dotted line). The colored lines represent the energy losses of a few individual electrons. 

In Figure~\ref{fig: 5Myr 100 GeV}, we also show the electron energy evolution over 5~Myr, for an initial energy of 100~GeV, showing that the effects on stochastic inverse-Compton scattering are also relevant at GeV-scale energies, where Klein-Nishina suppression is weak. The average final stochastic energy is 32~GeV for an initial energy of 100~GeV with +/-5~GeV at $1\sigma$, and the final energy in the analytic calculation is 32~GeV. The average number of ICS interaction is 1692 in the 100-GeV case, which is very close to the 1961 interactions in the 10~TeV case (Figure~\ref{fig: electron energy vs time}). The datasets contain 1000 particles.

We note that this last result is quite important -- electrons with an initial energy of 100~GeV are well within the Thomson regime for interactions with the dominant ISRF contributions from the CMB, IR emission and optical emission. They only barely lie in the Klein-Nishina regime for interactions with UV photons. Still, we find a significant ($>$15\%) dispersion in the final energy of the electron population after 5~Myr, which is slightly larger than the dispersion for our 10~TeV electrons. This strongly demonstrates that the phenomenon that we demonstrate is not limited to the Klein-Nishina regime. It is instead an important effect whenever the individual photon energy lost per ICS interaction is relatively large, and the number of individual ICS interactions is relatively small -- an effect that is still true at energies near 100~GeV. \\

\section{Isotropic Inverse-Compton Scattering} \label{app: isotropic}
In many scenarios, the ISRF is isotropic, based on contributions from many optical and infrared sources. Thus, many studies use an isotropic version of the ICS cross-section, which is equivalent to Equation~\ref{eq:fullkn} integrated over solid angle:

\begin{multline}\label{eq: isotropic cross section}
\frac{dN\left(E_\gamma\right)}{dE_\gamma} = \frac{2\pi r_0^2}{\nu_i E^2} \times  \left[1 + \frac{z^2}{2\left(1-z\right)} + \frac{z}{b\left(1-z\right)} - \right. \\
\left. \frac{2z^2}{b^2\left(1-z\right)^2} - \frac{z^3}{2b\left(1-z\right)^2} -\frac{2z}{b\left(1-z\right)}\ln{\left(\frac{b\left(1-z\right)}{z}\right)}\right],
\end{multline}

\noindent where $\nu_i$ is the initial photon energy, $E_\gamma$ the outgoing gamma energy, $E$ the electron energy before the interaction, and $z \equiv E_\gamma/E$ and $b \equiv 4\nu_iE$.

Figure~\ref{fig: electron energy vs time isotropic} shows the electron energy as a function of pulsar age for electrons produced at pulsar birth, 1~Myr ago, with an initial energy of 10~TeV for 1000 electrons. This is identical to Figure~\ref{fig: electron energy vs time} but for an isotropic ICS calculation. 
The analytic ICS calculation is shown as a black-dashed line, the average of the stochastic ICS as the black-solid line with the $1\sigma$ and $2\sigma$ bands in dark gray and light gray, respectively. The colored lines represent a few examples of individual electrons. The final energy in the exact stochastic ICS calculation is 287~GeV with an energy spread between 241--333~GeV at $1\sigma$ and 195--379~GeV at $2\sigma$, which is similar to the results of the non-isotropic case, as expected.  \\

\section{Treatment of Diffusion} \label{app: diffusion}
Throughout the bulk of this paper, we have focused on the effect of electron cooling on the total electron spectrum generated by a nearby, middle-aged pulsar. However, the electrons produced by this source also must diffuse through the interstellar medium, and the treatment of particle diffusion may affect the electron spectrum observed at Earth. 

We note several methods for proving that diffusion does not affect the production (in the analytic approximation) or smearing (in the correct stochastic model) of the spectral feature that we discuss.

\begin{figure}
\centering
\includegraphics[width=0.48\textwidth]{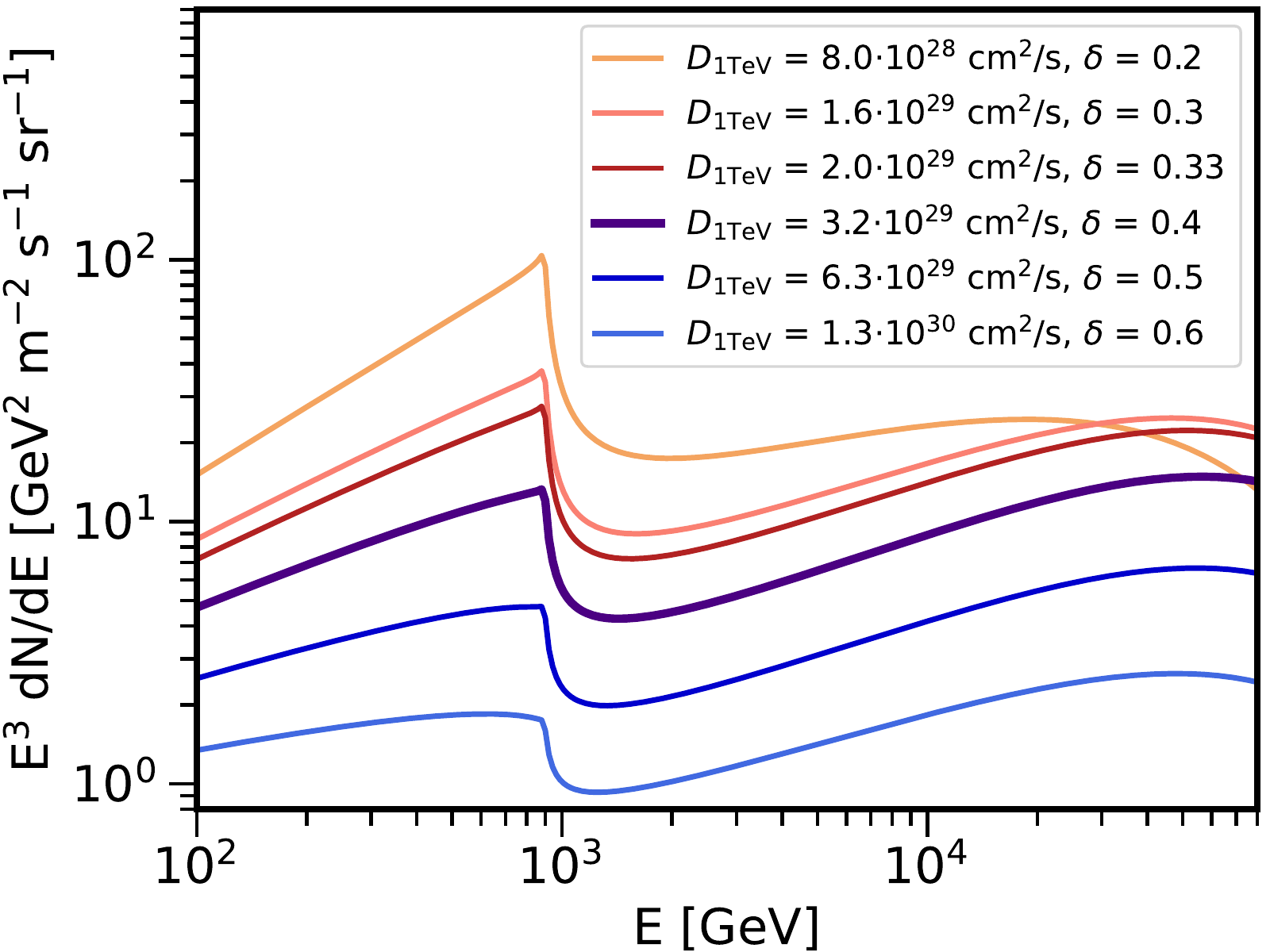}
\caption{The local electron flux observed at Earth calculated using the analytic approximation when diffusion is included. Results are shown for a number of diffusion indices and normalizations at a standard energy of 1~TeV. We note that both the location and sharpness of the spectral peak is not affected by the diffusion conditions, implying that it is generated by the cooling term, and is insensitive to particle diffusion.}
\label{fig: diffusion plot}
\end{figure}

First, we note that diffusion does not affect the production of the sharp spectral feature in the analytic approximation. In Figure~\ref{fig: diffusion plot} we show the local electron flux produced by Geminga using multiple different choices for the diffusion coefficient (at 1~TeV) and the diffusion spectral index. Because this changes the energy-dependent fraction of cosmic-rays that are near the Earth position after 370~kyr, it changes both the normalization and the overall spectrum of the electron flux both below and above the peak. However, the location and sharpness of the spectral feature is mostly unaffected, because it stems purely from the effective cooling of very-high-energy electrons.

To show this, we directly create a version of our stochastic model that includes diffusion. This is computationally difficult, because the diffusion of each individual electron must be simulated via Monte Carlo techniques along with its energy losses. After the simulation is complete, electrons will be discarded unless they have diffused to the correct distance between the source and observer. This process cannot be separated, because the individual interactions of each electron influence their energy, and thus the efficiency through which they diffuse through the interstellar medium.

To produce this simulation, at each time step, we calculate the diffusion coefficient for an individual electron by:

\begin{equation}\label{eq: diff}
D = D_0 \left(\frac{E}{1\text{ GeV}}\right)^\delta,
\end{equation}
where $D_0$ is the normalization of the diffusion coefficient at 1~GeV and $\delta$ the diffusion spectral index. We adopt typical values of $D_0 = 2\times10^{28}$~cm$^2$/s and $\delta = 0.4$~\cite{Hooper:2017gtd}. Using the diffusion coefficient, the mean free path of the electrons can be calculated by $\delta r = 6D/c$, where $c$ is the speed of light. Since diffusion can be modelled as a 3-dimensional random walk, we choose a random direction in which the electron travels the calculated distance.

\begin{figure*}[tbp]
\begin{minipage}[t]{0.48\textwidth}
\centering
\includegraphics[width=0.98\textwidth]{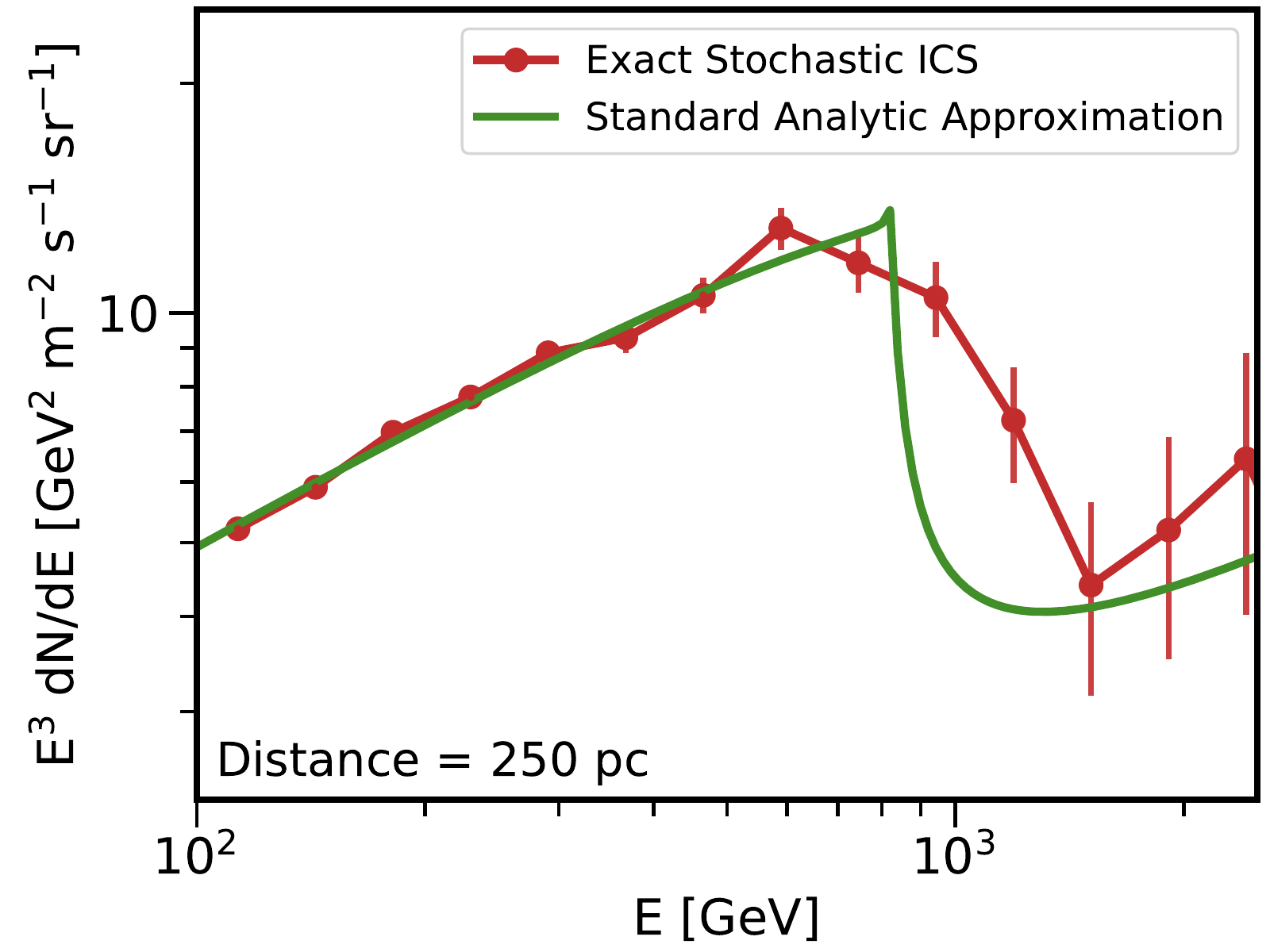}
\end{minipage}
\hfill
\begin{minipage}[t]{0.48\textwidth}
\centering
\includegraphics[width=0.98\textwidth]{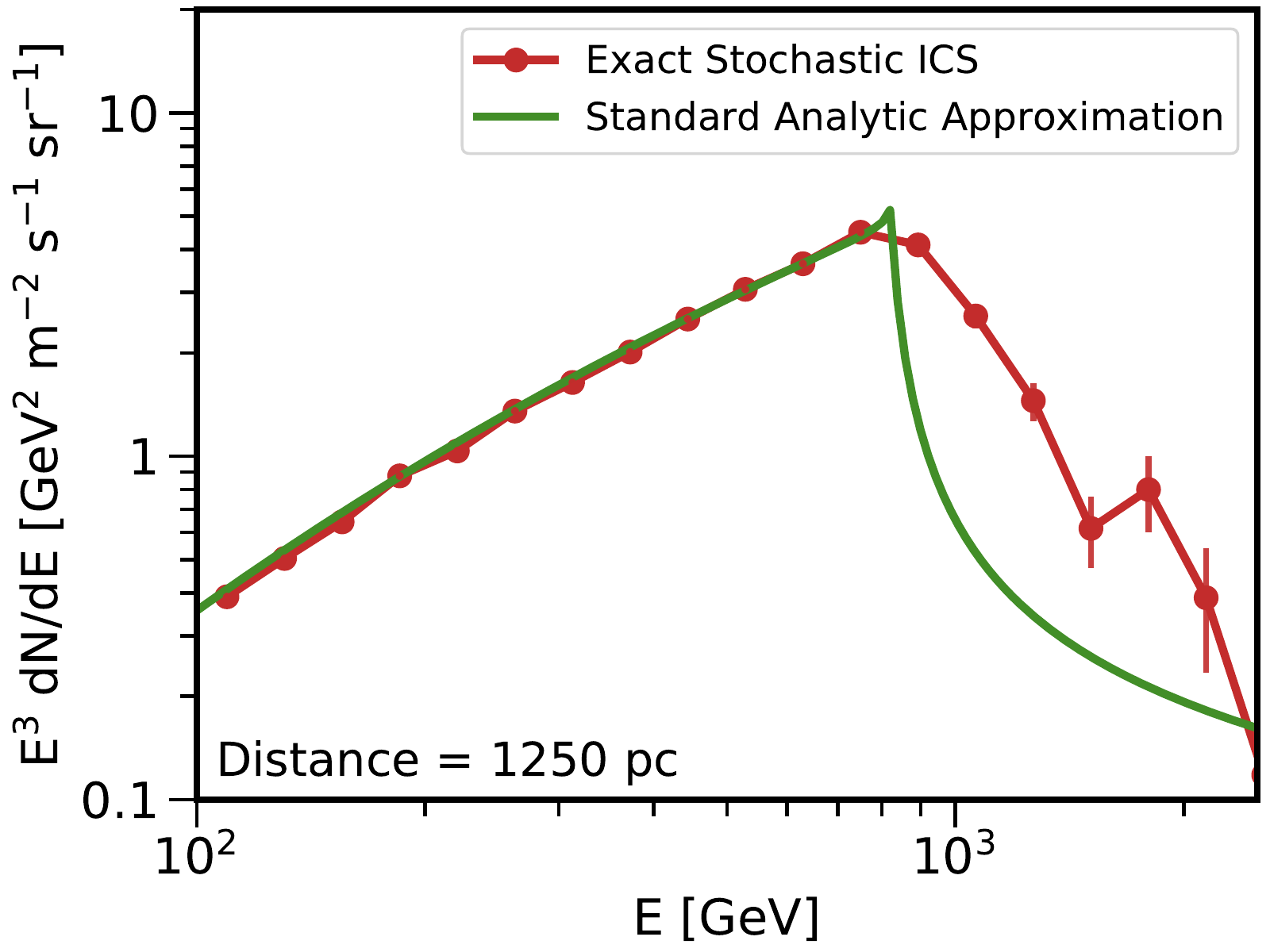}
\end{minipage}
\caption{Comparison of the analytic and stochastic inverse-Compton scattering models for a diffused spectrum. The left panel shows the spectrum at 250~pc, the distance from Geminga to Earth, while the right panel shows the spectrum at 1250~pc.}
\label{fig: propagation spectrum}
\end{figure*}

The spectrum for Geminga (at age 370~kyr) can be seen in Figure~\ref{fig: propagation spectrum}. The left panel shows the diffused spectrum at 250~pc for the analytic approximation and the stochastic model. For better statistics, we include all electrons whose final positions are within the range of 200 to 300~pc in the plot (some amount of binning is necessary because the probability that an individual electron lands exactly at Earth is miniscule even though this introduces a slight additional smearing form the binning). The average expected displacement of an electron with an initial energy of 1~TeV is given by $L_\text{diff} = \sqrt{6Dt}$, which gives $L_\text{diff} \approx 1500$~pc in 370~kyr. This means that most electrons end up at a distance much further away from Geminga than Earth. In the right panel, we show the diffused spectrum at 1250~pc, near the location where the total electron flux (integrated over the concentric ring) is maximized. Because the electron count is much larger at this distance than closer to the pulsar, we can obtain good statistics with a more limited radial range of 1225--1275~pc, which reduces any effect of smearing from our binning. 

In both cases, we clearly see an identical spectral effect as we produce in the main text (where diffusion is not considered). In each case there is a significant spectral cutoff in the analytic approximation, which stems from the fact that higher energy electrons are cooled to a common energy. In the stochastic modeling, this effect is smeared out due to the different energy loss histories of each individual electron. We note that the statistical uncertainties in each bin are slightly larger (especially in our model at 250~pc) purely due to the computational difficulties in simulating enough particles to produce a robust determination of the spectral feature.\\

\begin{figure*}[tbp]
\begin{minipage}[t]{0.48\textwidth}
\centering
\includegraphics[width=0.98\textwidth]{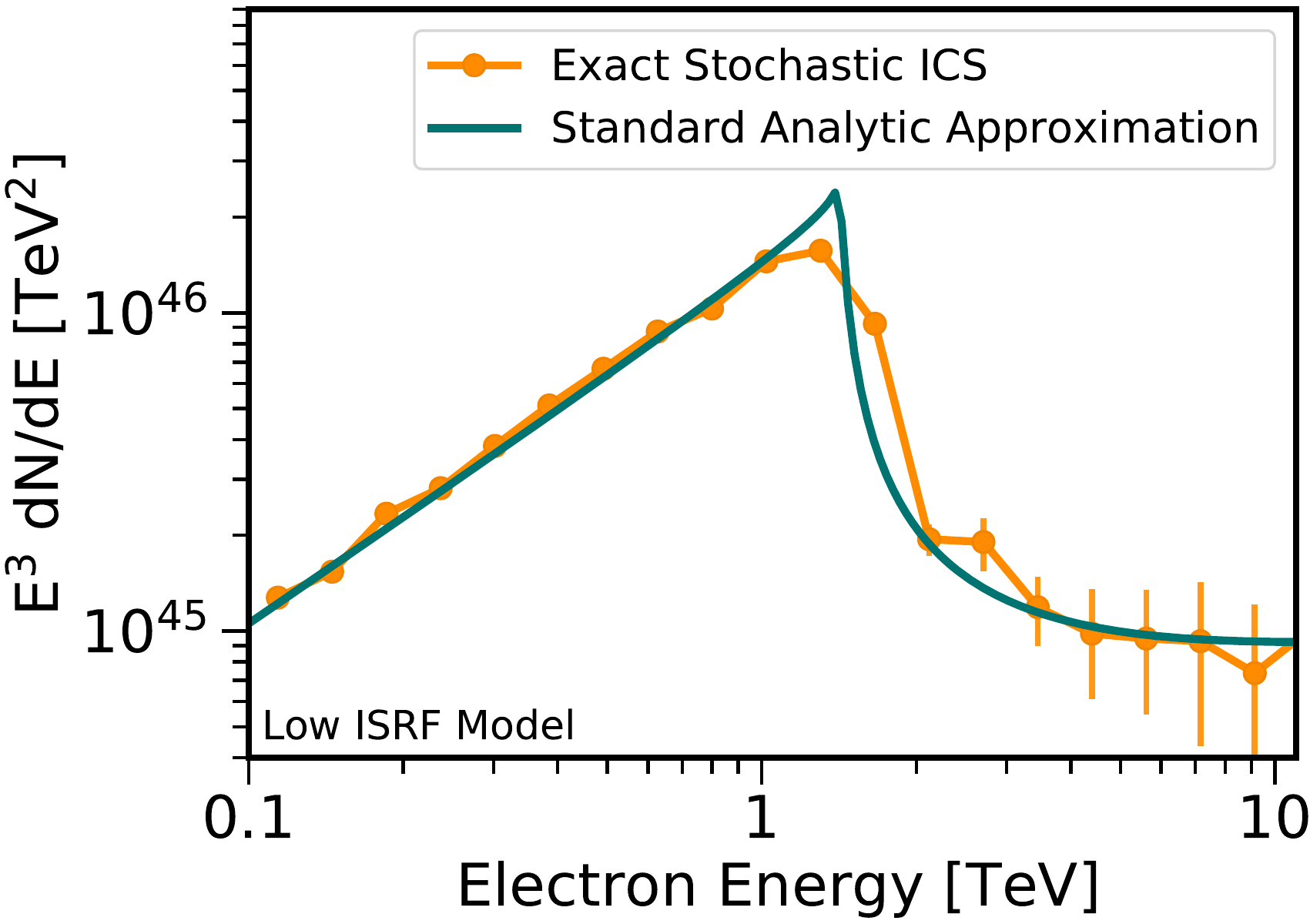}
\caption{Similar to Figure~1, but with an alternative ISRF model that has $\sim 30\%$ lower energy densities in the IR and optical component.}
\label{fig: fig1 low ISRF}
\end{minipage}
\hfill
\begin{minipage}[t]{0.48\textwidth}
\centering
\includegraphics[width=0.98\textwidth]{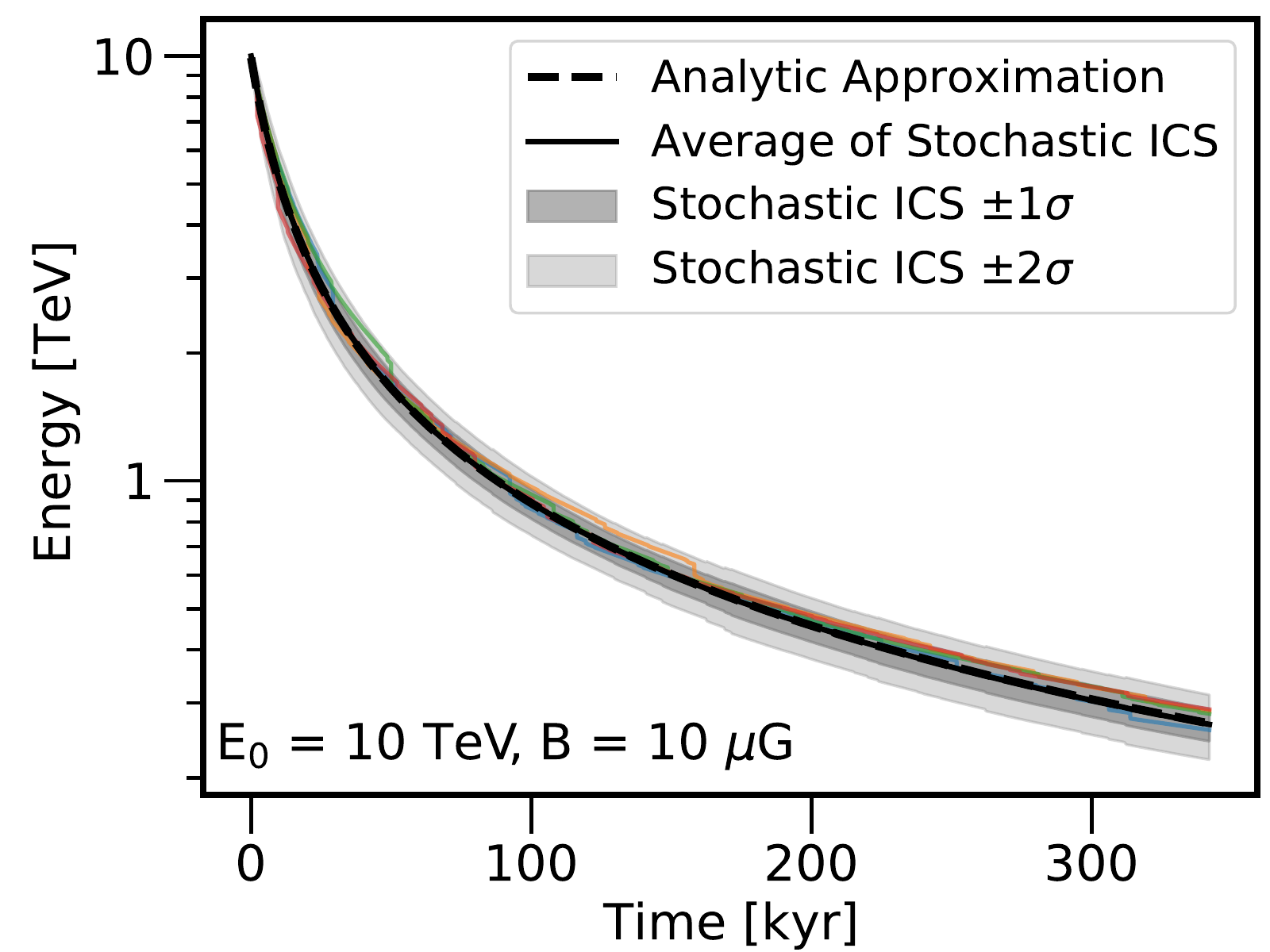}
\caption{Similar to Figure~3 in the main paper, but with an increased magnetic field strength of 10~$\mu$G instead of 3~$\mu$G, which means that synchrotron losses are at least 1.7 times stronger than the total ISRF losses. The average final stochastic energy is 267~GeV with an energy spread of +/- 23~GeV at $1\sigma$, and the final energy of the analytic model of 269~GeV.}
\label{fig: 10muG}
\end{minipage}
\end{figure*}

\section{Alternative Interstellar Radiation Field} \label{app: alternative ISRF}
Throughout our analysis, we adopt the interstellar radiation field model with the energy densities and photon temperatures as described in the main text: with temperatures \mbox{$T_\text{UV} = 20\times10^3$~K}, \mbox{$T_\text{optical} = 5\times10^3$~K}, \mbox{$T_\text{IR} = 20$~K} and \mbox{$T_\text{CMB} = 2.7$~K}, and  energy densities \mbox{$\rho_\text{UV} = 0.1$~eV/cm$^3$}, \mbox{$\rho_\text{optical} = 0.6$~eV/cm$^3$}, \mbox{$\rho_\text{IR} = 0.6$~eV/cm$^3$}, \mbox{$\rho_\text{CMB} = 0.26$~eV/cm$^3$}~\cite{Hooper:2017gtd}. 

However, the precise values for the ISRF are not well-known. To study the effect of the underlying ISRF model on our results, we employ an ISRF model with lower energy densities, and re-create Figure~\ref{fig:1} in the main text. We lower the energy densities of IR and optical radiation from 0.6~eV/cm$^3$ to 0.2~eV/cm$^3$, while keeping everything else the same. 

In Figure~\ref{fig: fig1 low ISRF}, we show the total flux at 342~kyr, for a dataset of $\sim$ 10\,000 particles. We find that the energy of the spectral peak has increased in both the analytic approximation and the stochastic model, which is expected because we have decreased the ISRF and thus the electron cooling rate. However, the effect of this model on the spectral peak is not changed -- the analytic model still produces a sharp spectral feature (even though we have decreased the importance of ICS cooling compared to synchrotron cooling), while the feature is still eliminated in our stochastic modeling. \\

\begin{figure*}[tbp]
\begin{minipage}[t]{0.48\textwidth}
\centering
\includegraphics[width=0.98\textwidth]{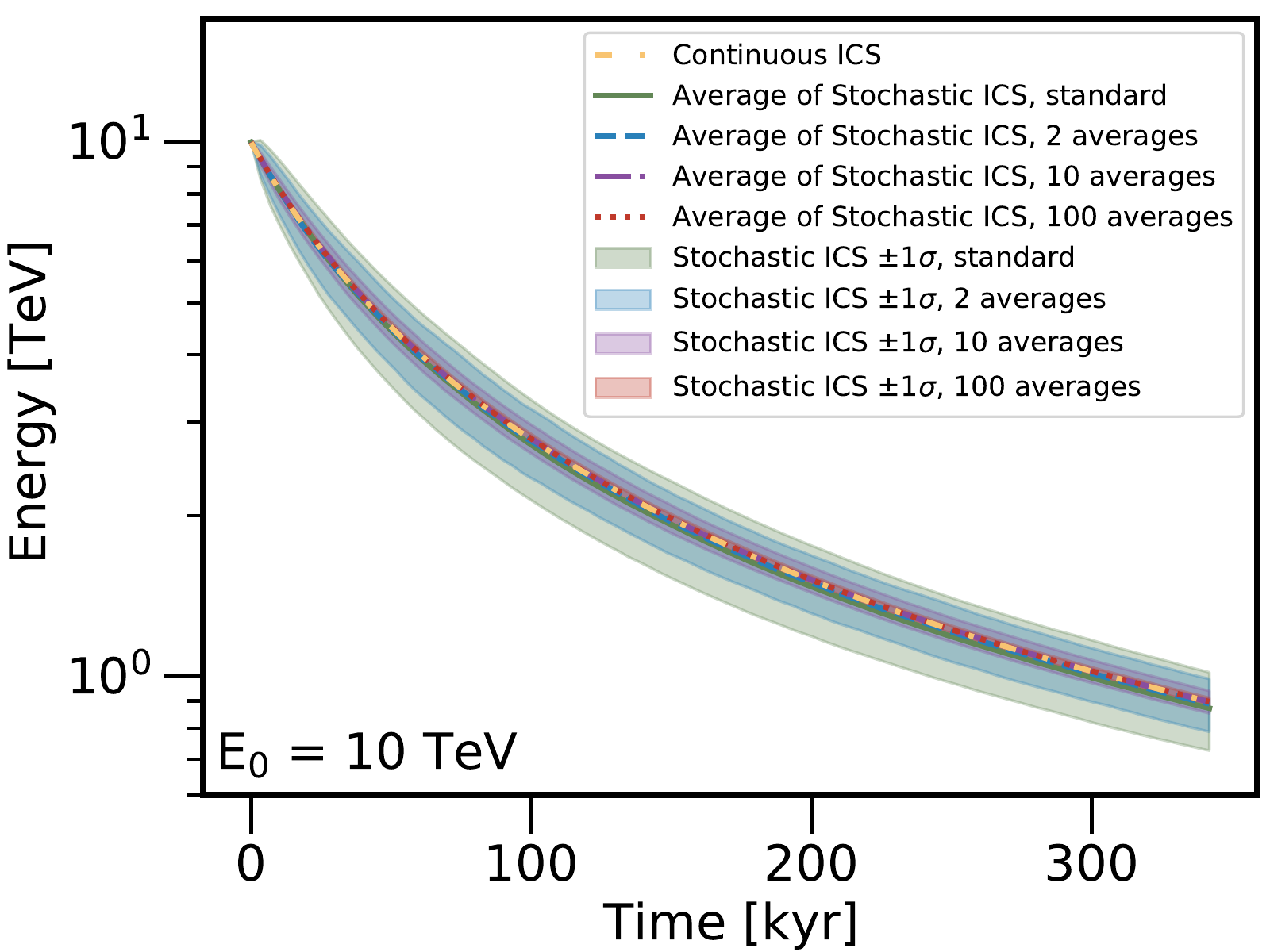}
\end{minipage}
\hfill
\begin{minipage}[t]{0.48\textwidth}
\includegraphics[width=0.98\textwidth]{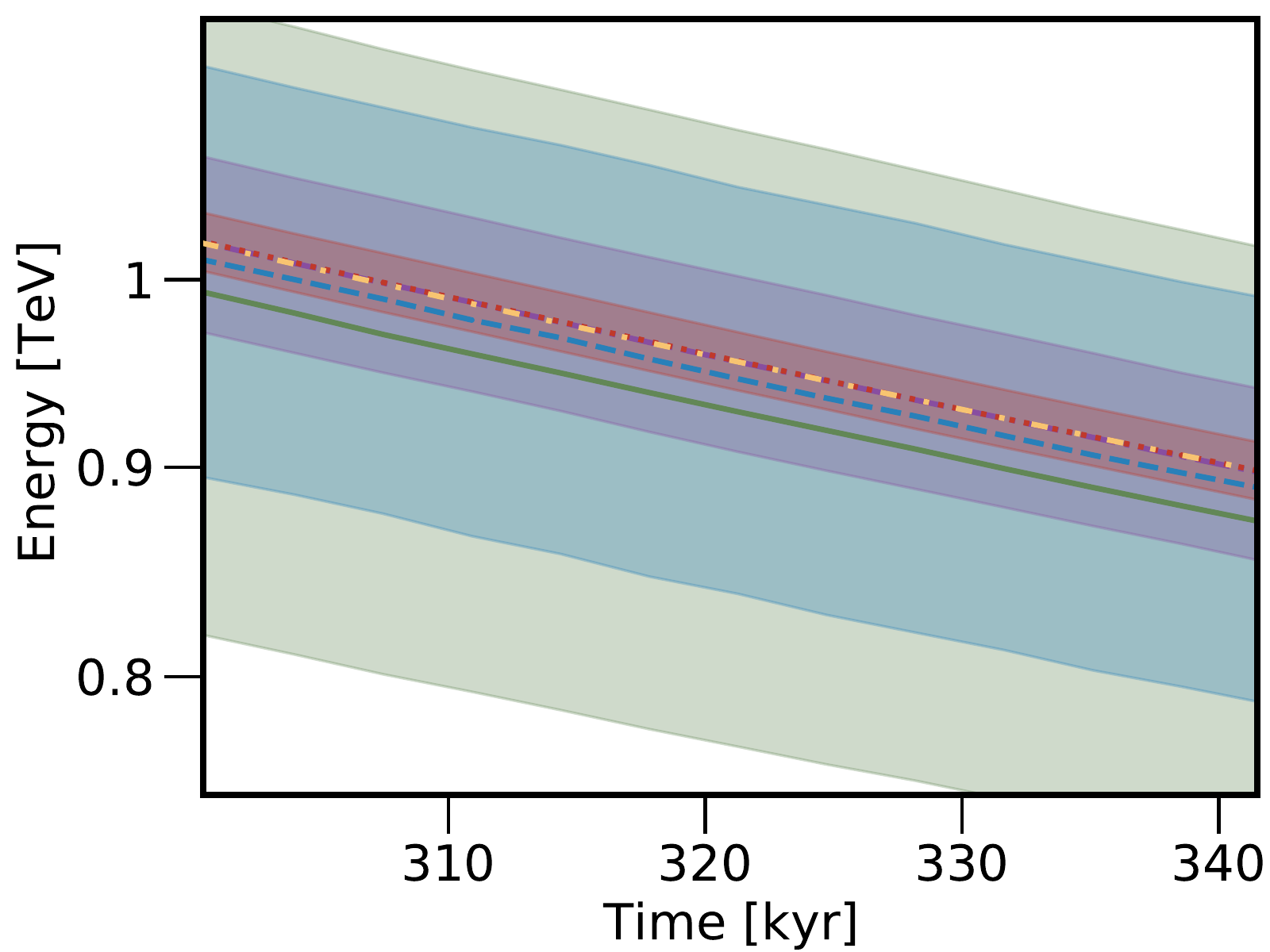}
\end{minipage}
\caption{Comparison of the standard stochastic average, and an average of 2, 10 and 100 at each time step, compared to the result of the analytic approximation. It can be seen that averaging the stochastic results reproduces the analytic result, which is already achieved after about 10 averages. The left panel shows the energy losses over 342~kyr, while the right panel shows a zoomed-in version for 300 to 342~kyr.}
\label{fig:averages}
\end{figure*}

\section{Alternative Magnetic Field Model} \label{app: strong magnetic field}
Throughout our analysis, we assume a Galactic magnetic field strength of $B = 3\,\mu$G, which determines the energy losses due to synchrotron radiation. This means that energy losses are dominated by ICS interactions up to electron energies of about 40~TeV (see the main text for the ISRF model), which includes most of the total electron power injected in our pulsar model.

Here we investigate the case where energy losses due to synchrotron radiation exceed ICS energy losses, and show that the stochasticity stemming from the limited number of ICS signal does not go away. So long as ICS losses are non-negligible -- particles will continue to be dispersed in energy due to the Poisson nature of these interactions. While keeping everything else in our model the same, we change the magnetic field strength to an extremely large value of 10~$\mu$G (which significantly exceeds current best-fit measurements). For this model, synchrotron losses dominate over ISRF losses for every electron energy in our simulation. We simulate energy losses for electrons with an initial energy of 10~TeV, similar to Figure~\ref{fig: electron energy vs time} for 342~kyr.

The result is shown in Figure~\ref{fig: 10muG}. We find that the average dispersion only decreases from 16\% (for $B=3\,\mu$G) to 8\%. Moreover, we note that -- due to the fact that the final electron energy has decreased due to the enhanced synchrotron losses -- the feature would actually be more detectable in current AMS-02 data, because the statistical precision of AMS-02 data is much better at lower energy. Thus, we argue that our results remain robust for all reasonable choices of the local magnetic field strength.

\section{Stochastic vs Continuous Loss Rates} \label{app: stochastics vs continuous loss rates}
Throughout this work (e.g., Figures~\ref{fig: electron energy vs time}, \ref{fig: electron energy vs time 3 TeV}, \ref{fig: 5Myr 100 GeV}, \ref{fig: electron energy vs time isotropic}), we note an interesting feature. The continuous loss-rate obtained from the analytic calculation is slightly larger ($\sim$2\%) than the average energy obtained from the stochastic model. While this effect is small, it is intuitively unexpected, because the analytic model (which essentially calculates the average effect from a scenario where every electron continuously interacts with every photon in the model), appears like it should provide a reasonable calculation for the average energy of an electron after a given time.

In fact, we find that the small difference between the stochastic and continuous energy loss rates stems from the energy-dependence of the inverse-Compton scattering cross section. In the stochastic scenario, the probability of having significant energy losses makes certain electron energies (and thus, certain inverse-Compton scattering cross-sections), more likely than other values. For example, a single interaction of a 10~TeV electron may remove many TeV of its energy, bringing it to a lower energy where the Klein-Nishina suppression with regards to the infrared portions of the ISRF are no longer as large. This slightly changes the expected energy of the electron after a time $t$ compared to a continuous energy loss model, where every electron moves through every energy value between the initial and final value. Since the ICS cross-section changes with energy, this leads to slightly different effective cooling rates.

To verify that this is the case (and that there are not any underlying issues with either our analytic approximation or stochastic model) we make a simple adjustment in our stochastic model. In each propagation time step, instead of using Monte Carlo techniques to select a single random input photon energy and final $\gamma$-ray energy, we wrap the Monte Carlo process in a for-loop and repeat the energy loss calculation several times. We then calculate the average energy loss for the Monte Carlo draws in our for-loop, and apply the average energy loss to our electron before moving onto the next time step.

Through the addition of a single for-loop, this allows our Monte Carlo code to continuously vary between the stochastic energy loss model (when the for-loop is run a single time), and the continuous energy-loss model (as the for loop is run infinite times). Specifically, in the limit of an infinite loop, every electron has interactions with every photon in the ISRF, and loses a small, but non-zero amount of energy in every time step, even if the time steps become very small. 

In Figure~\ref{fig:averages} we show the energy losses against time for 342~kyr for different number of repetitions of our loop at each time step. In the standard case, the energy loss is calculated once per time step, which is exactly our stochastic model. In the other cases, the energy losses are repeatedly drawn and averaged 2, 10 and 100 times at each time step, which begins to approach a continuous limit. The left panel shows the energy losses of the full 342~kyr of electron cooling, while the right panel shows the same but zoomed in to the last few 10~kyr. It can be seen that, for an increasing loop-length, the final electron energy after 342~kyr continuously changes from the average value calculated by the stochastic energy loss model, to the value calculated by the continuous energy loss model. Specifically, at an age of 342~kyr, the final average stochastic energies are 872~GeV in the standard case, 889~GeV for 2 averages, 896~GeV for 10 averages, and 897~GeV for 100 averages. The continuous loss rate, which purely relies on the standard analytical calculation as described in the main text, produces a final energy of 897~GeV, exactly matching our averaged (over 100 draws) stochastic code.

\begin{figure}[tbp]
\centering
\includegraphics[width=0.48\textwidth]{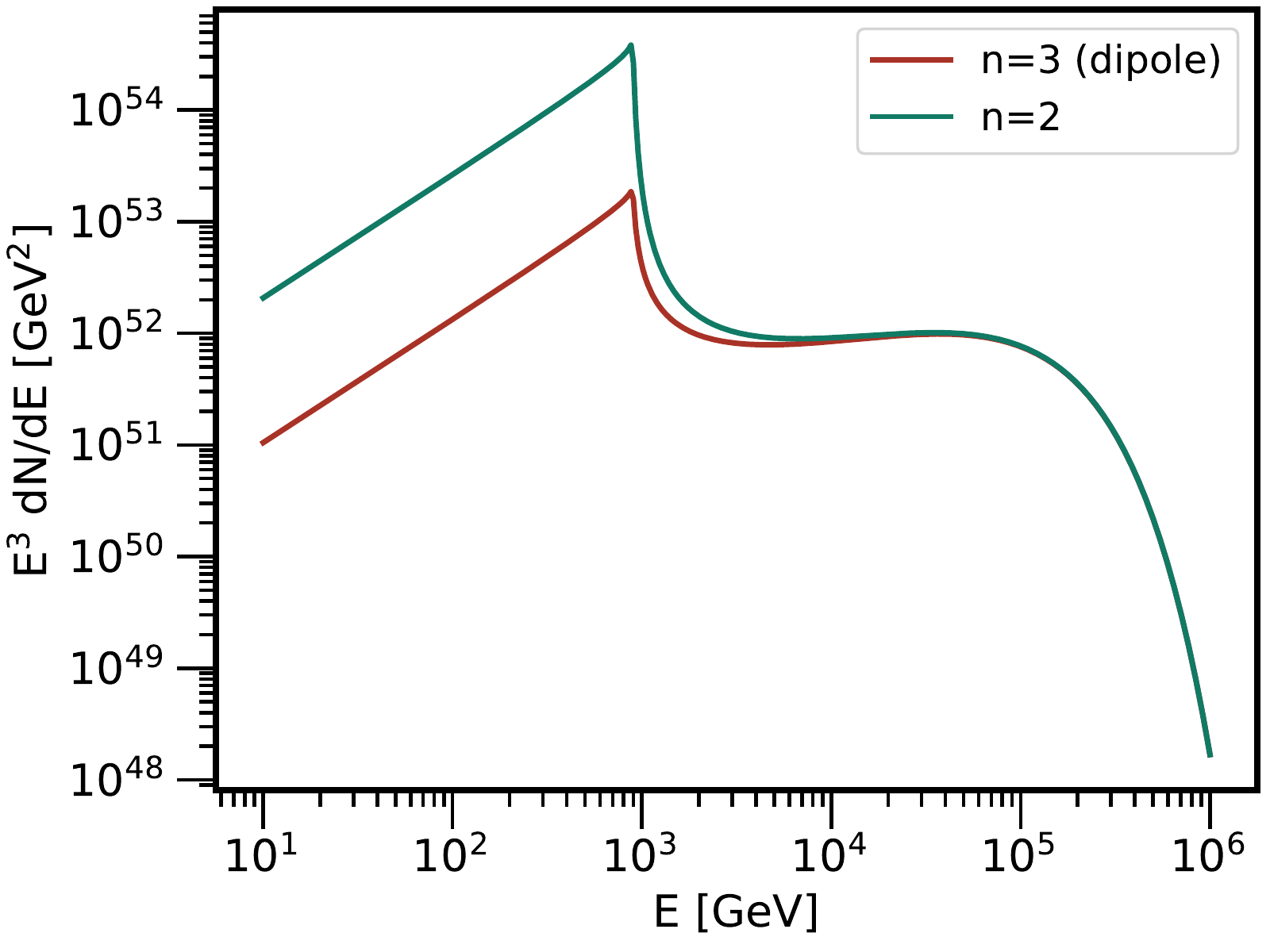}
\caption{Comparison of the total flux after 342~kyr for Geminga, assuming a dipole model ($n = 3$), which we have adopted throughout this work, and a non-dipole model with a braking index of $n = 2$. Since a smaller braking index increases the electron injection at early times, the sharp cutoff feature is enhanced when the braking index is lowered compared to the dipole model.}
\label{fig: dipole}
\end{figure}

We note that while this effect is extremely interesting -- it is practically undetectable because it is degenerate with the exact values of the magnetic field strength, the amplitude of the ISRF, and the pulsar age. If all of these parameters were known to within 2\%, then conceivably the error in the calculation of the average electron energy could be observed. This differs from the dispersion in the electron energies, which we show is robust for many different pulsar inputs -- and is already potentially detectable with existing AMS-02 data.

\section{Non-Dipole Pulsar Models} \label{app: non-dipole}
In Equation~\ref{eq:luminosityvstime}, we assume the braking index of the pulsar to be $n = 3$, which corresponds to a dipole model. However, observations suggest that pulsars are not an exact dipole and the braking index is lower, $n \sim 1.4 - 2.9$~(e.g.~\cite{Xu:2001bp, Lyne:2014qqa, Hamil:2015hqa}). Here, we study the effect of a smaller braking index on the pulsar feature. In Figure~\ref{fig: dipole}, we show the total flux from the analytic calculation after 342~kyr, in the dipole model with $n = 3$, that we assume throughout this work, and a non-dipole model with $n = 2$. In the non-dipole case, the sharp spectral feature becomes even more pronounced because more electrons are injected at earlier times compared to the dipole model.

We note that even a sharper spectral feature will be washed out in our correct stochastic modeling. This is apparent because we have shown in the main text, that even a delta-function injection signal (at 10~TeV) does not produce a spectral bump once stochastic energy losses are taken into account.

\bibliography{main}

\end{document}